\documentclass[prb,amsmath,amssymb, notitlepage, twocolumn,
nofootinbib,
superscriptaddress
]{revtex4-1}
\usepackage{amsmath,amssymb,bm,mathrsfs,graphicx, braket, times,MnSymbol}
\usepackage[colorlinks=true,citecolor=blue,linkcolor=blue]{hyperref}
\usepackage{longtable}
\usepackage{multirow}
\usepackage[all,cmtip]{xy}
\usepackage[normalem]{ulem}
\usepackage[usenames,dvipsnames]{color}
\usepackage{multirow}

\renewcommand{\vec}[1]{\bm{#1}}

\newcommand{\moire} {moir{\' e} }

\begin{document}
\title{
Faithful Tight-binding Models and Fragile Topology of Magic-angle Bilayer Graphene
}
\author{Hoi Chun Po }
\affiliation{Department of Physics, Harvard University,
Cambridge, MA 02138, USA}
\affiliation{Department of Physics, Massachusetts Institute of Technology,
Cambridge, MA 02139, USA}
\author{Liujun Zou}
\affiliation{Department of Physics, Harvard University,
Cambridge, MA 02138, USA}
\affiliation{Department of Physics, Massachusetts Institute of Technology,
Cambridge, MA 02139, USA}

\author{ T. Senthil}
\affiliation{Department of Physics, Massachusetts Institute of Technology,
Cambridge, MA 02139, USA}

 \author{Ashvin Vishwanath}
 \affiliation{Department of Physics, Harvard University,
Cambridge, MA 02138, USA}

\begin{abstract}
Correlated insulators and superconductivity have been observed in ``magic-angle'' twisted bilayer graphene, when the nearly flat bands close to neutrality are partially filled.
 While a momentum-space continuum model accurately describes these flat bands, interaction effects are more conveniently incorporated in tight-binding models. We have previously shown that no fully symmetric tight-binding model can be  minimal, in the sense of capturing just the flat bands, so  extended models are unavoidable. Here, we introduce a family of tight-binding models that capture the flat bands while simultaneously retaining all symmetries. In particular, we construct three concrete models with five, six, or ten bands per valley and per spin. These models are also faithful, in that the additional degrees of freedom  represent energy bands further away from neutrality, and they serve as optimal starting points for a controlled study of interaction effects. Furthermore, our construction demonstrates the ``fragile topology'' of the nearly flat bands; i.e., the obstruction to constructing exponentially localized Wannier functions can be resolved when a particular set of trivial bands is added to the model.
\end{abstract}

\maketitle

\section{Introduction}
In strongly correlated materials such as transition metal oxides, which include the high-$T_c$ cuprate materials \cite{Lee2006}, the competition between kinetic energy and electron-electron interactions stabilizes remarkable phases such as Mott insulators and high-temperature superconductors. Their theoretical description traditionally begins with a tight-binding model which provides a real-space representation of the relevant electronic bands. Interactions are then incorporated by means of a local $U$ term, leading to the Hubbard model.

Recently, another example of a correlated insulator in proximity to a superconductor has appeared---two adjacent graphene sheets that are twisted by a specific small angle relative to each other \cite{Cao2018, Cao2018a}. Here, we will address the question of constructing a minimal model for twisted bilayer graphene,  analogous to the square-lattice tight-binding model for cuprates. We will see that traditional approaches to this problem fail due to a form of band topology inherited from the underlying Dirac nature of the problem. Instead, a different approach is called for, which is developed in this paper.

Twisted bilayer graphene (TBG) structures have been studied intensely in the last decade \cite{Neto2007, STMNatPhy, Magaud2010, Mele2010, Shallcross2010,  Bistritzer2010, Morell2010,  STMPRL, Bistritzer2011, CastroNeto2011,Castro-Neto2012, He2013, Uchida2014, Yan2012, Jung2014, Sboychakov2015, Crommie2015,  ShiangPRB, PabloPRL, Kim2017, Nam2017,  LeRoy2018, Ensslin2018, Ramires}.
  To begin with, the two valleys of graphene are decoupled from one another, particularly in the limit of small twist angles, yielding a valley quantum number in addition to spin. The electronic states near each valley of each graphene monolayer hybridize with the corresponding states from the other monolayer.  When the twist angle is close to certain discrete values known as the magic angles, e.g., at $\sim 1.05^\circ$, theoretical calculations show that there are two nearly flat bands (per valley and per spin) that form in the middle of the spectrum \cite{Bistritzer2011} and are separated from other bands \cite{Nam2017}. 
The band gaps are also observed in experiments \cite{Cao2018, Cao2018a}.
These nearly flat bands contain Dirac nodes that intersect the chemical potential at neutrality.

Counting electron filling from neutrality, at  fillings $\nu_T = \pm 4$ a band insulator is obtained. However, in experiments, correlated insulators are observed at partial band fillings of $\nu_T = \pm 2$ on cooling below a few degree Kelvin. Further doping this insulator at $\nu_T=-2$ with either electrons or holes reveals superconductivity at a $T_c \sim 1$ K \cite{Cao2018a}. Two natural questions that arise are the following: how similar is this phenomenon to that in the cuprates, and, relatedly, what is the minimal model that we should consider here, analogous to the  square-lattice Hubbard model for the cuprates?

Ideally, a minimal tight-binding model for TBG would describe {\em only} the two nearly flat bands (per valley and per spin) and respect all symmetries.
However, there is an interesting topological aspect to the nearly flat bands that obstructs finding such a model \cite{Mar2018,Jun2018}. This is intuitively seen by recognizing that the two Dirac points in the nearly flat bands originate from the unperturbed Dirac cones which, while living in different layers, belong to the same valley. This suggests that they carry the same chirality \cite{CastroNeto2011, He2013, Goerbig2017, Cao2018}, i.e., the same Berry phase  of $\pm \pi$ of the two Dirac cones, the relative sign of which is well defined. In contrast, any two-band tight-binding model would give Dirac cones with vanishing net chirality. This obstruction is also reflected in the symmetry eigenvalues of the bands, which cannot be captured in a minimal two-band tight-binding model \cite{Mar2018,Jun2018}. Given that a minimal tight binding model is forbidden, one must then proceed with one of the following options (i) extend the model to include additional bands,  (ii) give up on some symmetries; or (iii) construct effective orbitals which are not exponentially localized in all spatial directions \cite{Z2Wannier,Wannier-Qi, WannierSheets}.

In this paper we will pursue the first option, namely, forgoing the minimality requirement and constructing models with more than two bands.
Specifically, we will introduce a ten-band model
in which the connected bands follow a 4-2-4 sequence, with the middle two bands representing the nearly flat bands of a single valley of TBG. Our model has the following advantages. First, all symmetries are respected and represented appropriately, and the two isolated bands incorporate the previously mentioned band topology of TBG. Second, the additional, complementary bands have a natural correspondence with the higher-energy bands in TBG. Finally, we can incorporate an approximate particle-hole symmetry in this description, which is known to be a good symmetry for the higher-energy states (although generally broken for the nearly flat bands). We also discuss the construction of more minimal models with  six or five bands which retain a set excited bands only on one side of the nearly flat bands.  Our models therefore pave the way to the derivation of a symmetric, interacting real-space description of TBG. Our solutions are  reminiscent of `p-d' models of the copper oxides, where correlated `d' copper orbitals are augmented by oxygen `p' orbitals \cite{Matsumoto1989}.
An important difference here is that the topological obstruction prevents further downfolding that would eliminate the additional bands.

Furthermore, the complementary bands in our model are manifestly {\it trivial}, in that they can be smoothly deformed into an explicit atomic insulator. This is conceptually interesting, as it is in stark contrast with the familiar forms of band topology, say those exemplified by the Haldane \cite{Haldane} or Kane-Mele \cite{Z2_QSH, Z2_Graphene} models, which are described by stable (K-theoretic) topological invariants \cite{Kitaev, FreedMoore,TeoKane}. There, the nontrivial topological indices must cancel when all sets of bands are accounted for, so a band with stable topology cannot be neutralized by additional trivial bands. Rather, our model proves that the identified band topology in TBG falls into the class of  ``fragile topology''  recently introduced in Ref.\ \onlinecite{Fragile}.
We stress that this identification of the fragile nature of the band topology in TBG has important implication in the construction of realistic tight-binding models. Suppose, in contrast, that the band topology was conventional. Then, to construct faithful tight-binding models for TBG, one must first identity the topological counterpart of the active bands among the the high-energy bands, similar to how the conduction bands in both the Haldane and Kane-Mele models are also topological. In contrast, our result on the fragile nature of the band topology in TBG implies one can construct effective tight-binding models simply by disentangling (in the sense of Ref.\ \onlinecite{VanderbiltRMP}) a suitable set of atomic states out of the high-energy degrees of freedom in TBG.

Let us also mention that, in Ref.\ \onlinecite{Jun2018}, we provided a different recipe to extend the model. 
There, we constructed a four-band model where all the symmetries are implemented naturally, and the four bands split into 2-2 sets of isolated bands, each of which individually showcases the identified band topology of TBG.
While the smaller number of bands is an advantage,  the additional bands are disconnected from the physical degrees of freedom in TBG, and interactions can be reliably treated only when they are weak enough that interband mixing can be safely neglected . In contrast, we believe our present solution is superior in that the additional degrees of freedom correspond to physical excited states, and that it clarifies the fragile nature of the band topology.

The second option for constructing effective tight-binding models, which we recall by way of review, is to circumvent the Wannier obstructions by implementing some of the symmetries in a nontrivial manner \cite{Mar2018}. This was done for valley symmetry in Ref.\ \cite{Mar2018}, which, however, entails a non-standard procedure to eventually recover the symmetry, unlike the option discussed here. Alternately, one can simply ignore some symmetries in the problem \cite{NoahLiang,Oskar, KoshinoLiang}, or adopt some different implementations of the symmetries \cite{XuBalents,  Dodaro2018, Zhang2018, Rademaker2018,Thomson2018}. 
An unintended consequence is a need for fine tuning. Hence, the theoretical predictions of these models are not automatically justified. For instance, in the symmetry setting of Refs.\ \onlinecite{NoahLiang,Oskar, KoshinoLiang} a vertical electric field would lead to a band gap at charge neutrality, which is inconsistent with that dictated by the actual symmetries of the system \cite{Neto2007}.

We begin by briefly reviewing the symmetries and band topology relevant for small-angle TBG, before defining tight-binding models and providing a physical picture for their construction. Finally we discuss the fragile topology of the TBG flat bands and close with a discussion.
Henceforth, we will focus on the single-valley problem with spin ignored.

\section{Symmetry and topology of TBG}
In the following, we focus on the symmetries of the continuum description \cite{Neto2007, Bistritzer2011}, which are also exact for the highest-symmetry commensurate lattice realizations \cite{Mele2010, Shallcross2010, Castro-Neto2012, Jun2018}. A more thorough review of these topics can be found in Ref.\ \onlinecite{Jun2018}.

The spatial symmetry group of TBG is generated by lattice translations, $C_6$ rotation, and a 2D mirror $M_y$ which flips the $y$ coordinate (more accurately, a layer-exchanging two-fold rotation in 3D)\footnote{
Note that $M_y$ may not be present in the actual experiment, due to, for instance, different processing of the top and bottom layers. Nonetheless, as long as its (explicit) breaking is perturbative, it will be helpful to retain it as a theoretical crutch.
}.
It is also symmetric under time-reversal $\mathcal T$. In addition, at small twist angle the two valleys are effectively decoupled, leading to an additional $U_{v}(1)$ valley charge conservation.
This allows one to consider the electronic degrees of freedom residing in a single valley.
Among the listed symmetries, only $C_6$ and $\mathcal T$ exchange the two valleys; all the others, as well as the combinations like $C_6 \mathcal T$, leave the valley charge invariant.
Consequentially, the (magnetic) point group of the single-valley problem is generated by  $C_6 \mathcal T$ and $M_y$, and the problem is described by the magnetic space group 183.188 (in the BNS notation) \cite{Mar2018}.

One can readily compute the symmetry representations furnished by the two nearly flat bands at different high-symmetry momenta, which we list in Table \ref{tab:SymRep}.
Importantly, one can check that {\it no} atomic insulator with the same symmetries will have the same set of symmetry representations (Appendix \ref{app:SymRep}), and therefore there is an obstruction for constructing symmetric Wannier functions for the two nearly flat bands. This alone implies the two relevant bands are topologically obstructed from any tight-binding description that respects all symmetries \cite{Mar2018}.
In addition, unlike the familiar case of monolayer graphene \cite{Goerbig2017}, the two Dirac points in the TBG band structures have the {\it same} chirality \cite{CastroNeto2011, He2013, Goerbig2017, Cao2018}. This is impossible in any two-band tight-binding model and constitutes another Wannier obstruction \cite{Mar2018}. Curiously, the two mentioned obstructions, derived respectively from the representations of $M_y$ and the net chirality of the Dirac points, are intertwined: it was shown in Ref.\ \onlinecite{Jun2018} that when the only Dirac points are pinned to the two (moir{\'e}) K points, the mirror and chirality obstructions are equivalent.

\begin{center}
\begin{table}
\caption{Symmetry representations furnished by the nearly flat bands of twisted bilayer graphene \cite{Mar2018, Jun2018}, indicated by the eigenvalues of the generating symmetries of the point group.
Eigenvalues from degenerate bands are grouped by parenthesis.
\label{tab:SymRep}}
\begin{tabular}{c|cc}
\hline \hline
Eig.\ & $\Gamma$ & K\\
\hline
$C_3$ & $1$, $1$ & $(\omega, \omega^*)$\\
$M_y$ & $1$, $-1$ & not a symmetry\\
\hline \hline
\end{tabular}
\end{table}
\end{center}

\section{Tight-binding models}
In this section, we introduce three tight-binding models involving different number of bands, namely, ten, six or five, which are summarized in Table \ref{tab:TBSummary}. Note that, throughout the paper, the band counting assumes one focuses on a single valley of TBG with spin ignored. 
As will be evident later, all the models we present are constructed in the same spirit: in each of the models, there will be two groups of bands. The first group comprises two bands around zero energy, which faithfully captures all the symmetry, topology and energetic features of the active bands in TBG; the second group, which we will call ``complementary bands,'' corresponds to the higher-energy bands in TBG. We will later show in Sec.\ \ref{sec:Fragile} that the complementary bands in the models we construct are all topologically trivial, in the sense that the full filling of these complementary bands gives rise to an atomic insulator. 
Despite their trivial nature, these complementary bands cannot be discarded from the model, as their presence is essential for resolving the topological obstruction in any symmetric, real-space description involving only the active bands \cite{Mar2018, Jun2018}.

\begin{center}
\begin{table*}
\caption{Summary of fully symmetric tight-binding models which capture the key features of the two active bands in twisted bilayer graphene.
\label{tab:TBSummary}}
\begin{tabular}{c|cccc}
\hline \hline
~~~Number of bands~~~
 & 
~~~\multirow{2}{*}{Complementary bands are atomic}~~~&
~~~Captures representations  ~~~&
~~~Possesses approximate  ~~~& 
~~~\multirow{2}{*}{Reference}~~~ \\
(per valley and per spin) & ~& of the higher-energy bands & particle-hole symmetry \\
\hline
4& $\times$ & $\times$  & $\times$ & Ref.\ \onlinecite{Jun2018} \\
5& $\checkmark$  & $\times$  & $\times$ & Appendix \ref{app:5Band} \\
6 & $\checkmark$  & $\checkmark$  & $\times$ & Appendix \ref{app:6Band} \\
10& $\checkmark$  & $\checkmark$  & $\checkmark$ & Appendix \ref{app:10Band} \\
\hline \hline
\end{tabular}
\end{table*}
\end{center}

\subsection{A ten-band model}
Let us now describe the mentioned ten-band model.
In our present symmetry setting, one can label the orbitals as being either $s$, $p_z$, or $p_\pm$.
Both $s$ and $p_z$ orbitals transform trivially under a $C_3$ rotation, but $p_z$ flips sign under $M_y$ while $s$ does not\footnote{
Recall, our ``mirror'' $M_y$ is really a two-fold rotation in 3D.
}.
In contrast, the orbitals $p_\pm \equiv p_x \pm i p_y$ are exchanged under $M_y$ and form a doublet.
A more systematic tabulation of the symmetry properties of the orbitals can be found in Appendix \ref{app:SymRep}.
Our ten-band model comprises a $p_z$ orbital and a pair of $p_\pm$ orbitals localized to sites forming a triangular lattice ($\tau$), a $s$ orbital on the kagome lattice ($\kappa$), and a pair of $p_\pm$ orbitals forming a honeycomb lattice ($\eta$). For brevity, we will describe the orbital content using the notation $(\text{lattice}, \text{orbital})$, e.g., $(\tau, p_z)$ denotes the $p_z$ orbitals localized to the triangular site.
Similarly, we denote the associated fermion operator by $\hat \tau_{p_z}$.
The described degrees of freedom are tersely summarized in Table \ref{tab:TBModel} in this notation.

\begin{center}
\begin{table}
\caption{Orbital content of the ten-band model. $\tau$, $\kappa$, and $\eta$ respectively denote the triangular, kagome, and honeycomb sites. $s$, $p_z$, and $p_{\pm}$ denote different orbital characters.
One can also construct a more minimal six-band model using only the orbitals listed to the left of the double vertical line.
\label{tab:TBModel}}
\begin{tabular}{c|ccc||c}
\hline \hline
Orbitals &
$(\tau, p_z)$ & $(\tau, p_{\pm})$ & $(\kappa, s)$ & $(\eta, p_{\pm})$\\
\hline
No.\ of bands & 1 & 2 & 3 & 4\\
\hline \hline
\end{tabular}
\end{table}
\end{center}

A prerequisite for any tight-binding modeling of TBG is the capability of producing two isolated bands with the targeted momentum-space symmetry representations (Table \ref{tab:SymRep}). This is guaranteed in our model by the following representation-matching equation:
\begin{equation}\begin{split}\label{eq:RepMatch}
(\tau, p_z)  \oplus (\tau, p_{\pm}) \oplus (\kappa, s)
\overset{{\rm rep.}}{=}  (\eta, p_{\pm}) \oplus (\text{target}),
\end{split}\end{equation}
which one can verify using the comprehensive tabulation of the symmetry data in Appendix \ref{app:SymRep}.
The physical meaning of Eq.\ \eqref{eq:RepMatch} is that, representation-wise, it is possible to construct an atomic insulator, with the same symmetry properties as $(\eta, p_\pm)$, within the six-band sub-Hilbert space defined by the content on the left-hand side. In our ten-band model, we add another, explicit set of $(\eta, p_\pm)$ orbitals to both sides of the above equation, so that the full representation-matching equation reads
\begin{equation}
\begin{split}\label{eq:RepMatch2}
&(\tau, p_z)  \oplus (\tau, p_{\pm}) \oplus (\kappa, s)\oplus (\eta, p_\pm)_0
\overset{{\rm rep.}}{=}\\
&\quad\quad\quad\quad\quad\quad
(\eta, p_{\pm})_1 \oplus (\text{target})  \oplus (\eta, p_{\pm})_2.
\end{split}
\end{equation}
Note that we have added subscripts $0$-$2$ to clarify that they represent different sets of physical degrees of freedom despite sharing identical symmetry properties.

\begin{figure}[h]
\begin{center}
{\includegraphics[width=0.5 \textwidth]{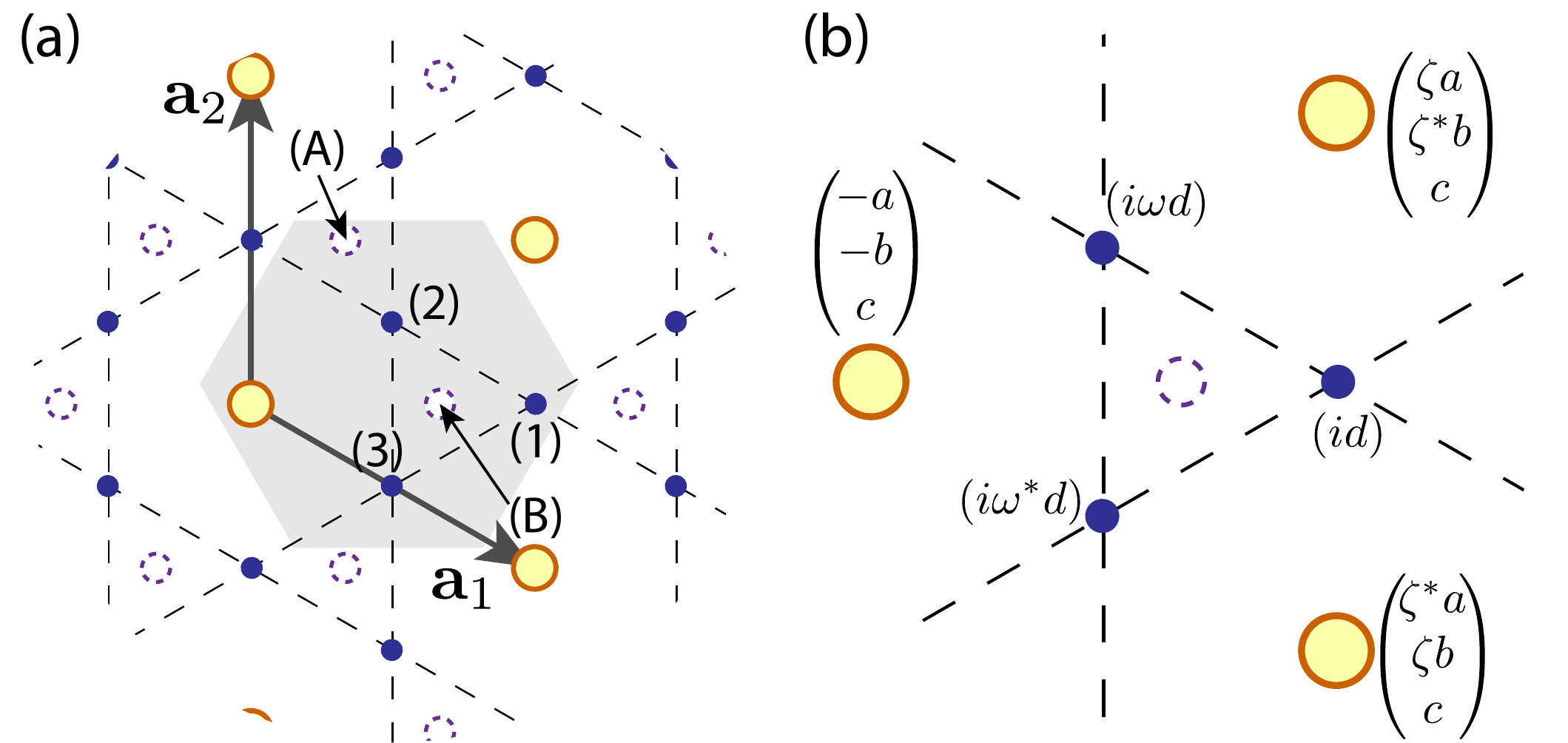}}
\caption{Real-space orbitals. (a) Lattices and conventions. The shaded region indicates the relative positions of the sites assigned to the same unit cell. (b) The constructed quasi-orbital $h_{p_+}^{(B)}$ in the real space. $\zeta = e^{i 2\pi/6}$ and  $\omega = \zeta^2 $.
Going from top to bottom, the entries in the three-component vectors attached to the triangular sites denote the amplitude for the $p_z$, $p_+$ and $p_-$ orbitals; those attached to the kagome sites denote the amplitude of their associated $s$ orbital.
\label{fig:Sites}
 }
\end{center}
\end{figure}

Guided by this observation, one can construct a model with the targeted representations simply through the construction of the mentioned atomic insulator.
Let $
\hat {\vec c}^\dagger_{\vec r}$ be the six-component fermion creation operators for the orbitals assigned to the unit cell at $\vec r$ (Fig.\ \ref{fig:Sites}a). We want to construct a localized ``quasi-orbital'' wave function $h_{p_+; \vec r}^{(l)} ( \vec x)$ such that $\hat h_{p_+;\vec r}^{(l) \dagger} \equiv \sum_{\vec x}\hat c^\dagger_{\vec x} h_{p_+;\vec r}^{(l)}( \vec x)$ has the same symmetry properties as a $p_+$ orbital centered at a {\it honeycomb site}, labeled by $l = A,B$ in the unit cell $\vec r$. This can be achieved by using a trial wave function which vanishes everywhere except on the three nearest kagome and triangular sites surrounding the honeycomb site.
Site symmetries reduce the freedom in the wave function to four real parameters\footnote{Or four complex parameters if we lift some fictitious symmetry constraints (Appendix \ref{app:Explicit}); one less if we also impose normalization.}, which we denote by $a$ through $d$ (Fig.\ \ref{fig:Sites}b).
Once $\hat h_{p_+;\vec r}^{(l) \dagger}$ is specified, using symmetries one can generate $\hat h_{p_-;\vec r}^{(l) \dagger}$ centered at the same site, as well as those centered on the other sites. Note that we have {\it not} imposed orthogonality between the $h_{p_\pm; \vec r}^{(l)}(\vec x)$ wave functions, and so their associated fermion operators do not obey the canonical anti-commutation relations.

We are now ready to define the ten-band model. Recall, in the above, we have not utilized the $(\eta, p_{\pm})_0$ orbitals in the system. Since they have identical symmetry properties as the $\hat h_{p_+;\vec r}^{(l) \dagger}$ quasi-orbitals we constructed, we can couple the two sets in a minimal manner:
\begin{equation}\begin{split}\label{eq:10Band}
\hat H(t,\delta)  = t \sum_{\vec r, l=A,B, \rho = p_\pm } \left(
\hat \eta^{(l)\dagger}_{\rho; \vec r } \hat h^{(l)}_{\rho; \vec r }
+  {\rm h.c.}
\right) + \delta \, \hat V,
\end{split}\end{equation}
where $t$ is a real parameter, $\hat V$ is a symmetry-allowed, local  perturbation which we detail in Appendix \ref{app:Explicit}, and the dimensionless parameter $\delta \in [0,1]$ controls the overall strength of the perturbation.
Note that the finite range of the wave functions $h_{p_+;\vec r}^{(l) \dagger}$ implies $\hat H$ is local\footnote{We remark that the present construction differs from the more conventional approach in which one tries to incorporate the smallest number of short-range hoppings to reproduce the salient features in the band structure. Instead, our goal here is to illustrate how the introduction of quasi-orbitals can naturally produce the nearly flat bands in TBG.}

.

\begin{figure}[h]
\begin{center}
{\includegraphics[width=0.49 \textwidth]{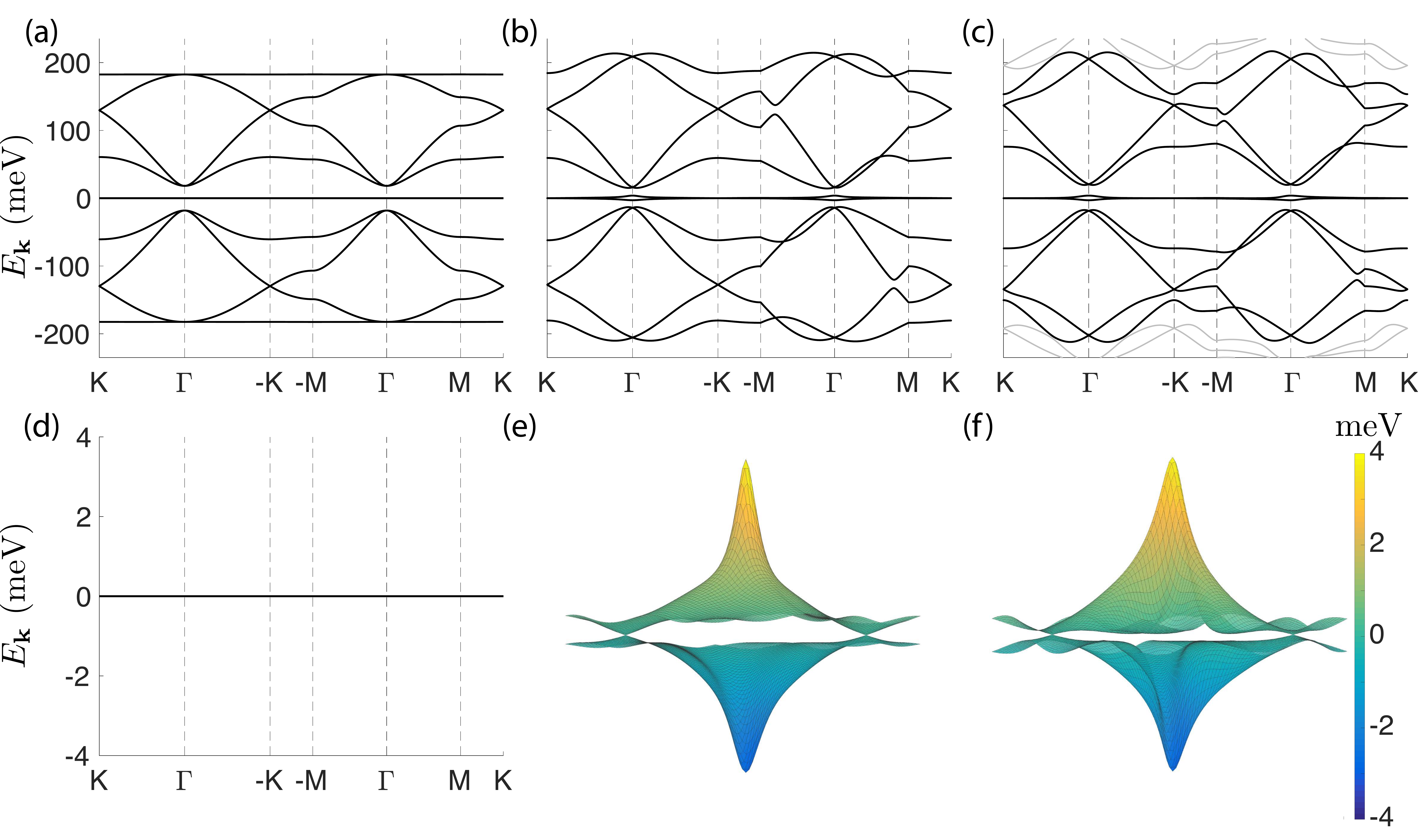}}
\caption{Band structures. (a,b) Bands from the ten-band Hamiltonian $\hat H(t_0,\delta)$.
For both panels, we choose $t_0 \equiv 130$ meV, and the wave-function parameters  $(a,b,c,d) = (0.110,0.033,0.033, 0.573)$.
We set $\delta = 0$ in (a) and $1$ in (b).
(c) Bands obtained from the continuum theory \cite{Neto2007, Bistritzer2011} for twisted bilayer graphene with a twist angle of $\theta = 1.05^\circ$, using the parameters described in Ref.\ \onlinecite{KoshinoLiang}. The ten bands around charge neutrality are highlighted. (d-f) A zoom-in of the two bands at charge neutrality for the corresponding panels in (a-c). The three-dimensional plots in (e,f) are plotted over the first Brillouin zone centered at $\Gamma$, showing the presence of exactly two Dirac points pinned to ${\rm K}$ and ${\rm K'}= {\rm -K}$. Note that (e) is generated from our tight-binding model, whereas (f) is generated from the continuum model.
\label{fig:Bands}
 }
\end{center}
\end{figure}

Although the perturbation $\delta \hat V$ in Eq.\ \eqref{eq:10Band} is needed for reproducing the detailed energetics features of the TBG band structures, we remark that the essential physics of the model can be understood by first setting $\delta = 0$, as is shown in Figs.\ \ref{fig:Bands}a. 
By construction, the band structure of $\hat H(t, 0)$ includes two exactly flat bands pinned at zero energy (Fig.\ \ref{fig:Bands}d; see also Appendix \ref{app:Explicit}), the symmetry representations of which must match those of the nearly flat bands in TBG. 
Very briefly, these flat bands exist here for the same geometric reason as that of the Lieb lattice \cite{Lieb}. To see why, consider any tight-binding model defined on a lattice with two sets of orbitals, which we label simply as $\alpha$ and $\beta$, and suppose that all the bonds connect an $\alpha$ orbital to a $\beta$ one. Due to this sublattice symmetry, the Bloch Hamiltonian automatically takes an off-diagonal form
\begin{equation}\begin{split}\label{eq:}
H_{\rm Lieb-like}(\vec k) = 
\left(
\begin{array}{cc}
0 & h_{\alpha \beta}(\vec k) \\
 h_{\alpha \beta}^\dagger (\vec k)& 0
\end{array}
\right).
\end{split}\end{equation}
Generally, the energy bands of such a Lieb-like Hamiltonian can be organized into bonding and anti-bonding pairs, which have energies $\pm E_{\vec k}$ due to the sublattice symmetry. However, if there are, say, more $\alpha$ orbital than $\beta$ ones, then the mentioned bonding/ anti-bonding picture cannot gap out all the $\alpha$ degrees of freedom from charge neutrality, which results in exactly flat bands pinned to $E=0$. More concretely, suppose there are $N_\alpha$ and $N_\beta$ orbitals in the two respective sets, then there will be at least $|N_\alpha - N_\beta|$ exactly flat bands in the spectrum of $H_{\rm Lieb-like}(\vec k)$ \footnote{
Slightly more technically, these flat bands arise because the rank of $h_{\alpha\beta} (\vec k)$ is at most $\min (N_\alpha, N_\beta)$. When $N_\alpha \neq N_\beta$, either the left or right null space of $h_{\alpha\beta} (\vec k)$ has to be at least $|N_\alpha - N_\beta|$ dimensional for all $\vec k$. This translates into zero-energy flat bands for $H_{\rm Lieb-like}(\vec k)$.
}.

Now recall that, in Eq.\ \eqref{eq:10Band}, the 	four ``quasi-orbtials'' $\hat h^{(l)}_{\rho; \vec r }$ are defined only using the degrees of freedom in the six bands on the left hand side of Eq.\ \eqref{eq:RepMatch}, whereas the additional orbitals $\hat \eta^{(l)}_{\rho; \vec r }$ are defined on an independent set of degrees of freedom (namely, $(\eta, p_\pm)_0$ on the left hand side of Eq.\ \eqref{eq:RepMatch2}).
As such $\hat H(t_0,\delta = 0)$ is Lieb-like. The mismatch between the number of degrees of freedom (six vs four) leads to two exactly flat bands pinned to charge neutrality.
In addition, we have chosen the wave function parameters $a$-$d$, listed in the caption of Fig.\ \ref{fig:Bands}, to reproduce the broad energetic features of the higher-energy bands of TBG. 
Note that our model reproduces the approximate $E_{\vec k} = -E_{-\vec k}$ particle-hole symmetry of the higher energy states in TBG, although this is not a good symmetry of the nearly flat bands.

With all the key properties built-in already, we simply choose $\hat V$ such that $\hat H(t_0,1)$ faithfully captures the energetics of the ten bands near charge neutrality in TBG.
This leads to the band structure shown in Fig.\ \ref{fig:Bands}b, which closely resembles that computed using the continuum theory of TBG (Fig.\ \ref{fig:Bands}c). In particular, the two bands near charge neutrality in Fig.\ \ref{fig:Bands}e touch only at the Dirac points pinned at K and K', just like those from the continuum theory (Fig.\ \ref{fig:Bands}f). As they furnish the targeted symmetry representations in Table \ref{tab:SymRep}, from the results in Ref.\ \onlinecite{Jun2018} they must display both the mirror and chirality Wannier obstructions, i.e., this ten-band model serves as an explicit resolution of all the known Wannier obstructions of the nearly flat bands of  TBG.

\subsection{A six-band model}
As the dominant term in the ten-band model in Eq.\ \eqref{eq:10Band} can be viewed  as a minimal coupling between the $\eta$ and $h$ degrees of freedom, one could imagine the consequences of ``integrating out'' the $\eta$ fermions, which results in a low-energy theory described in terms of the $h$ degrees of freedom. In our band-theory context, such a procedure can be done simply by adding an arbitrarily large chemical potential to $\eta$, which amounts to removing the four $\eta$ bands from the Hilbert space. The leads to a six-band low-energy Hilbert space with the orbital content on the left-hand side of Eq.\ \eqref{eq:RepMatch}, but with the dominant kinetic term involving only the four bands arising from the $h$ quasi-orbitals, i.e., there will again be two nearly flat bands near zero-energy.

\begin{figure}[h]
\begin{center}
{\includegraphics[width=0.49 \textwidth]{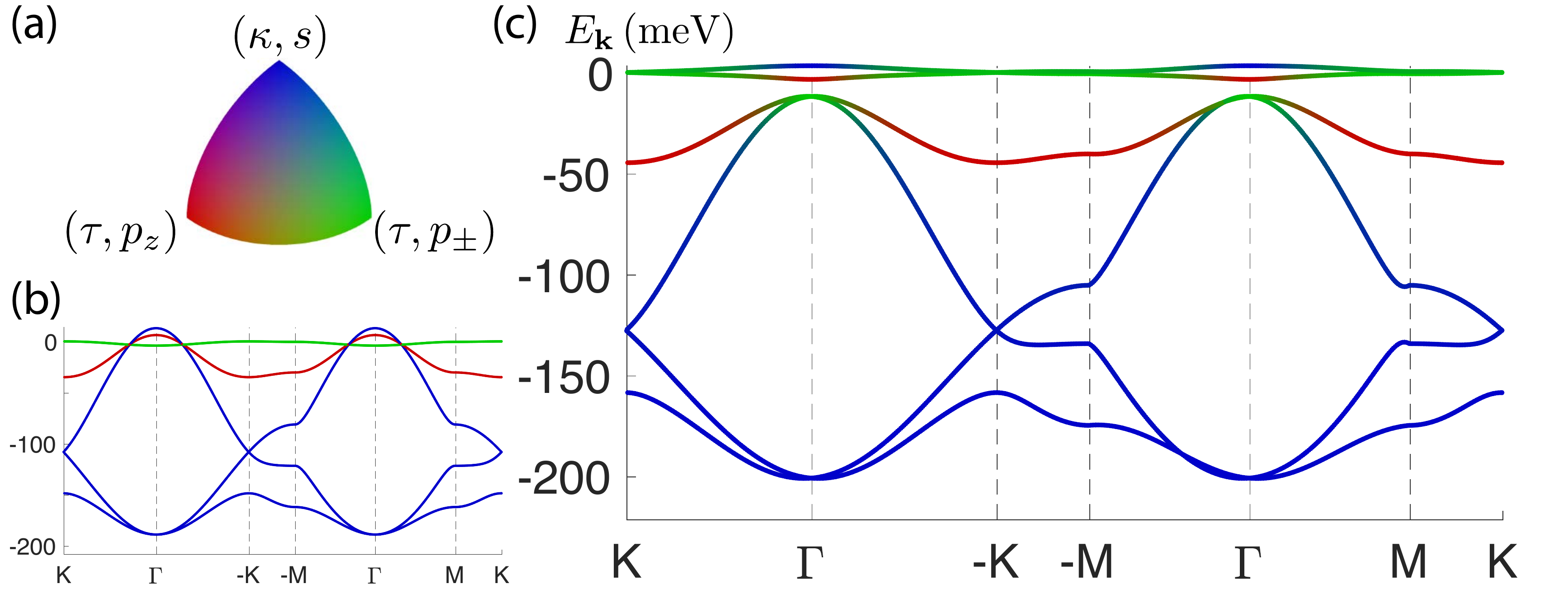}}
\caption{
Band structures from a six-band model.
(a) Color code for the orbital characters. (b) The broad energetic features can be set up using only the intra-orbital dispersion. (c) Band structure from the full model, with parameters detailed in Appendix \ref{app:6Band}.
\label{fig:6Band}
 }
\end{center}
\end{figure}

While the preceding picture explains the existence of a six-band model, it is also desirable to construct such a model in a more conventional manner in terms of mostly nearest neighbor bonds. 
We will undertake this task below.
Recall that the electron density of the nearly flat bands in TBG is localized to the ``AA'' regions, which  form a triangular lattice at the \moire scale \cite{Magaud2010, Uchida2014, PabloPRL, ShiangPRB,STMPRL, STMNatPhy,Crommie2015}. This suggests a tight-binding model with two orbitals placed on the triangular site. To capture the existence of Dirac points at K and K', these orbitals should be $p_\pm$, and naturally we anticipate the nearly flat bands to overlap strongly with the $(\tau, p_\pm)$ orbitals in most of the Brillouin zone. However, the $(\tau, p_\pm)$ bands feature an additional quadratic touching\footnote{
which could split into multiple Dirac cones when trigonal warping is incorporated
} at the Gamma point, which cancels the Dirac-point chirality.
In contrast, in TBG the two nearly flat bands are non-degenerate at $\Gamma$, and so the $(\tau, p_\pm)$ bands alone are incapable of capturing the $\Gamma$-point behavior \cite{NoahLiang, Mar2018}.
Therefore, we expect strong hybridization between the other orbitals in the vicinity of $\Gamma$, such that the wave function of the two nearly flat bands correspond to the singlet representations in $(\tau, p_z)$ and $(\kappa, s)$.

Based on the above picture, we construct a six-band model which captures all the salient features of TBG, as we show in Fig.\ \ref{fig:6Band} and elaborate on in Appendix \ref{app:6Band}.

\subsection{A five-band model
\label{sec:5Band}}
In the above, we have introduced two (closely related) tight-binding models, with ten or six bands, which captures the key properties of {\it both} the two nearly flat bands as well as the set(s) of four bands further away from charge neutrality. 
Yet, since our ultimate goal is to provide a real-space description of the two nearly flat bands, it might be beneficial to consider models with a smaller number of bands at the cost of a less accurate description of the high-energy states. 
In this subsection, we provide a five-band model constructed in this spirit.

We remark that we have already introduced a four-band model in Ref.\ \onlinecite{Jun2018}, which also captures the symmetry and band topology of the two active bands in TBG. However, in the four-band model the complementary bands have the same band topology as the active bands, and do not have the appropriate symmetry representations to reproduce the physical higher-energy bands in TBG. In contrast, our model here faithfully reproduces the features of the TBG bands in the energy window from roughly $-100$ meV to the top of the two nearly flat bands. 
As is evident below, this five-band model will still belong to the same class as those already constructed, and it will capture all the energetic, symmetry and topology features of the two nearly flat bands in TBG.
In addition, the three complementary bands in the model will again be atomic in nature. 

\begin{figure}[h]
\begin{center}
{\includegraphics[width=0.5 \textwidth]{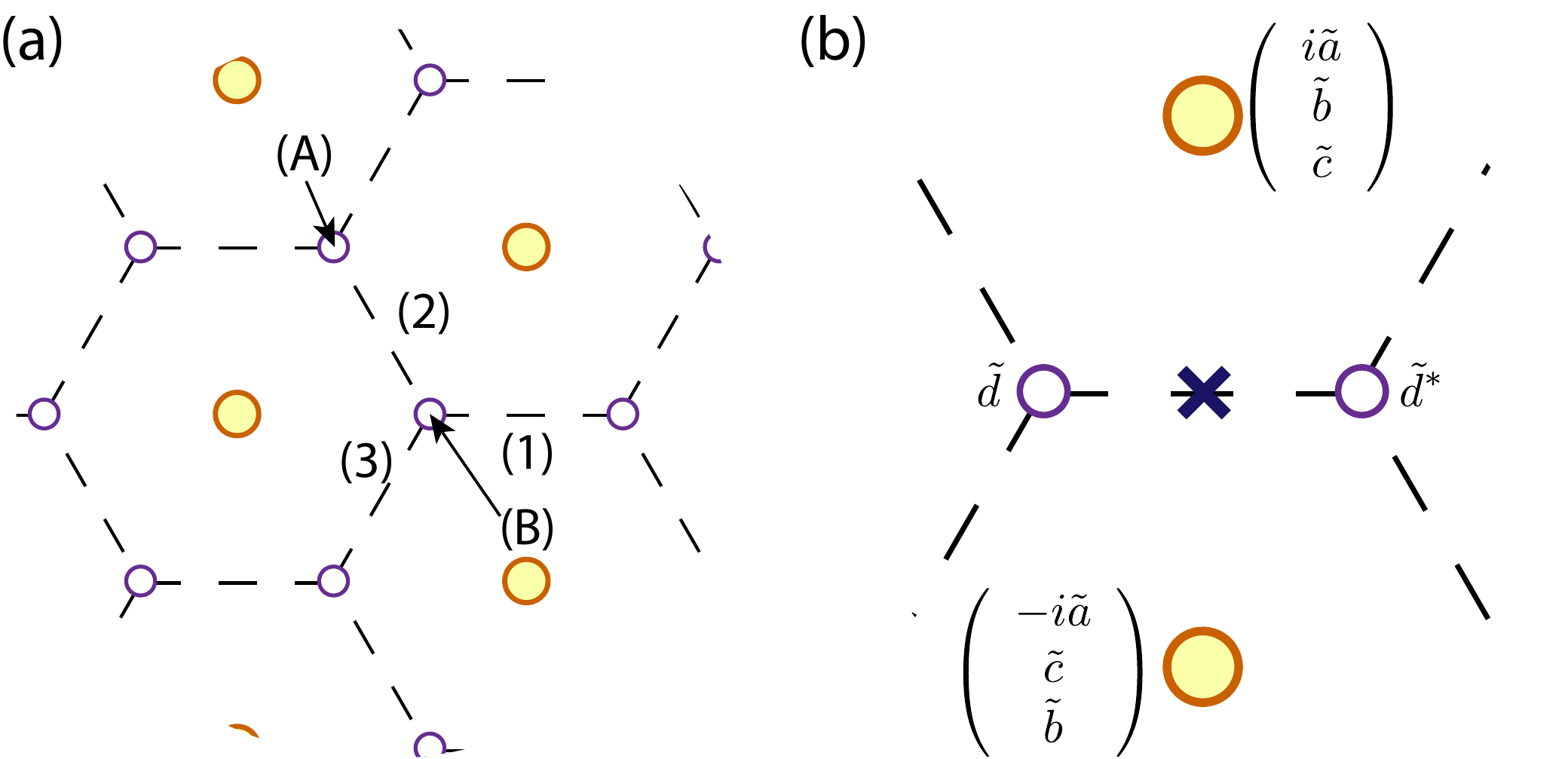}}
\caption{
Real-space orbitals for the five-band model. (a) On each of the triangular sites (filled yellow circles), we consider the three $p$ orbitals $p_z$ and $p_\pm \equiv p_x \pm i p_y$; on each of the honeycomb sites (open circles), we consider an $s$ orbital. The centers of the nearest-neighbor bonds between the honeycomb sites form a kagome lattice.
(b) The constructed quasi-orbital $\rho_{s}^{(1)}$ centered at a kagome site, indicated by a cross.
Going from top to bottom, the entries in the three-component vectors attached to the triangular sites denote the amplitude for the $p_z$, $p_+$ and $p_-$ orbitals; those attached to the honeycomb sites denote the amplitude of their associated $s$ orbital.
The other two quasi-orbitals labeled by $l=2,3$ can be obtained through symmetries.
\label{fig:Sites5}
 }
\end{center}
\end{figure}

In parallel with the preceding discussions, our starting point will be a representation-matching equation which is analogous to Eq.\ \eqref{eq:RepMatch}, but involves only five bands:
\begin{equation}\begin{split}\label{eq:5BandOrbitals}
(\tau, p_z) \oplus (\tau, p_\pm) \oplus (\eta, s)  \overset{{\rm rep.}}{=} (\kappa, s) \oplus ({\rm target}). 
\end{split}\end{equation}
We will again construct a set of pseudo-orbitals which have identical symmetry properties as $(\kappa, s)$ but residing in the Hilbert space defined by the five bands on the left. Due to a shortage of alphabets we will denote the quasi-orbital wave functions by $ \rho^{(l)}_{s;\vec r}$, where $l=1,2,3$ labels the three kagome sites in each unit cell. The site convention and a real-space description of  $\rho^{(1)}_{s; \vec r}$ are provided in Fig.\ \ref{fig:Sites5}. The other two wave functions can then be generated using the $C_3$ symmetry, and we have provided the explicit form of the Fourier transform of the wave functions in Appendix \ref{app:5Band}.

Next, we construct our Hamiltonian using the quasi-orbtials.
By design, when the three quasi-orbitals $ \rho^{(l)}_{s;\vec r}$ are projected away from zero energy, the remaining two bands will capture the essential properties of the active bands in TBG. Such projection can be effectively performed by simply giving these quasi-orbitals a negative ``on-site'' chemical potential $-t_0'$. 
Note that, given the nontrivial shape of these quasi-orbitals, such terms are not really on-site in the original degrees of freedom in the lattice, which are given by the left hand side of Eq.\ \eqref{eq:5BandOrbitals}; rather they should be viewed as specific hopping terms across the different orbitals.  We further introduce actual on-site potentials $\mu'_j$ as perturbation, which leads to the Hamiltonian
\begin{equation}\begin{split}\label{eq:5Band}
\hat H^{(5)} =  - t_0' \sum_{\vec r}  \sum_{l=1}^3 \hat \rho^{(l) \dagger}_{s;\vec r}\hat \rho^{(l)}_{s;\vec r}
+ \sum_{\vec r} \sum_{j}  \mu_j' \hat c_{j; \vec r}^\dagger  \hat c_{j; \vec r} ,
\end{split}\end{equation}
where $j$ runs over the five sites (per unit cell) in $(\eta, s) \oplus (\tau, p_z) \oplus (\tau, p_\pm)$.
The discussion on the corresponding Bloch Hamiltonian can be found in Appendix \ref{app:5Band}. We note that, similar to Eq.\ \eqref{eq:10Band}, the main term in Eq.\ \eqref{eq:5Band} serves to project the quasi-orbital degrees of freedom away from zero energy, which, by construction, leaves behind states that faithfully capture the nearly flat bands in TBG.

\begin{figure}[h]
\begin{center}
{\includegraphics[width=0.5 \textwidth]{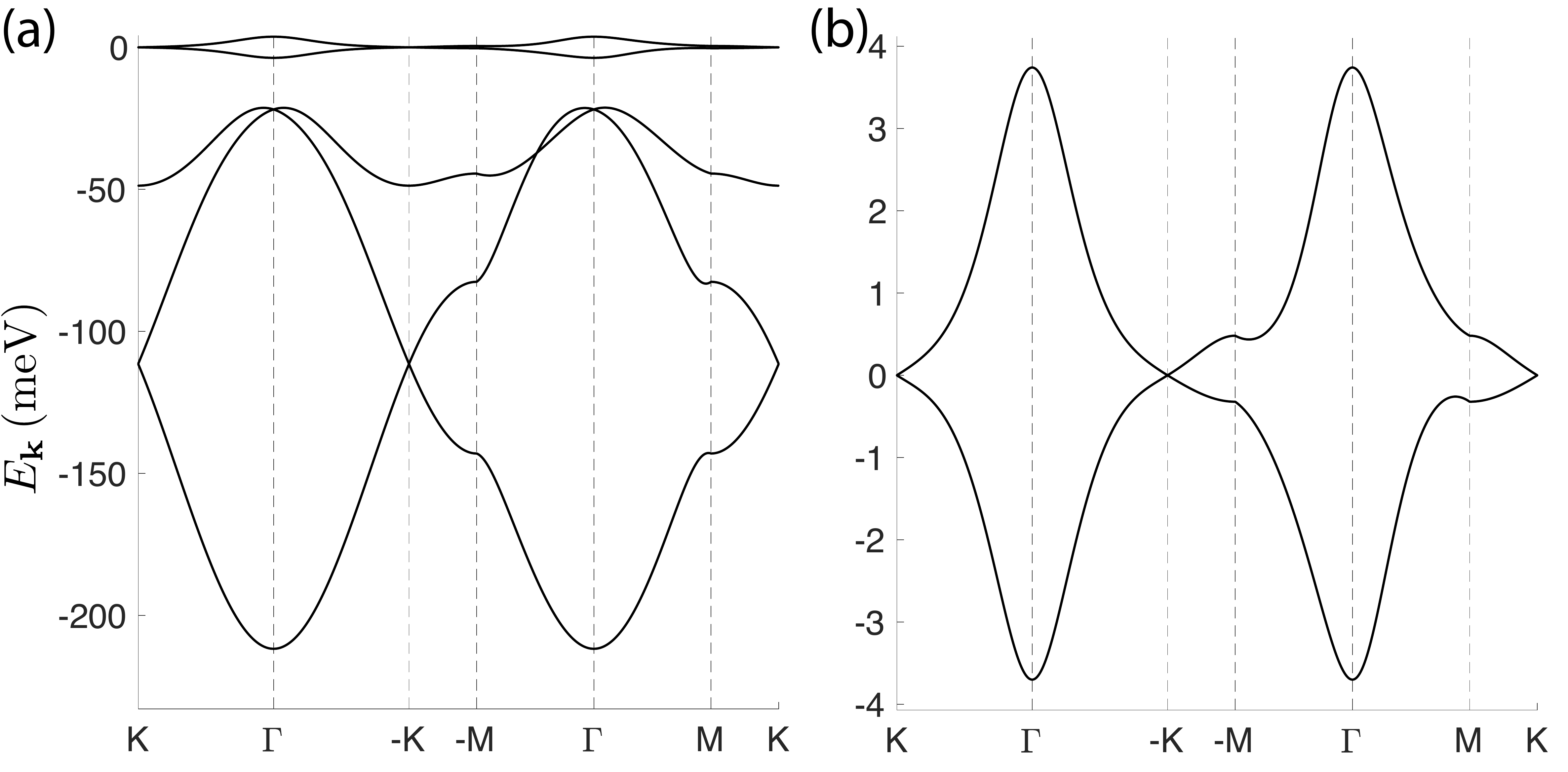}}
\caption{
Band structure from the five-band model. 
The following parameters are used: $(\tilde a, \tilde b, \tilde c, \tilde d) = (0.25, 0.2, 0.1, 0.67)$, $
( \mu_{p_z}' , \mu_{p_\pm}' , \mu_{\eta}') = (  -0.043 , 0, 0.05) t_0'$, and $t_0' = 80$ meV.
(a) The full spectrum. (b) A zoom-in for the two nearly flat bands.
\label{fig:5Bands}
 }
\end{center}
\end{figure}

The band structure of $\hat H^{(5)}$ is shown in Fig.\ \ref{fig:5Bands} (the parameters used are provided in the figure caption).
One sees that this five-band model faithfully captures the energetics from the top of the nearly flat bands down to $\sim -100$ meV. In addition, we again find exactly two Dirac points in the nearly flat bands, pinned respectively to K and K'. By design, all the symmetry representations of the nearly flat bands here are identical to those in TBG, and, therefore, based on the results in Ref.\ \onlinecite{Jun2018} we again conclude these two nearly flat bands showcase the known band topology of TBG.

\section{Fragile topology \label{sec:Fragile}}
\begin{figure}[h]
\begin{center}
{\includegraphics[width=0.48 \textwidth]{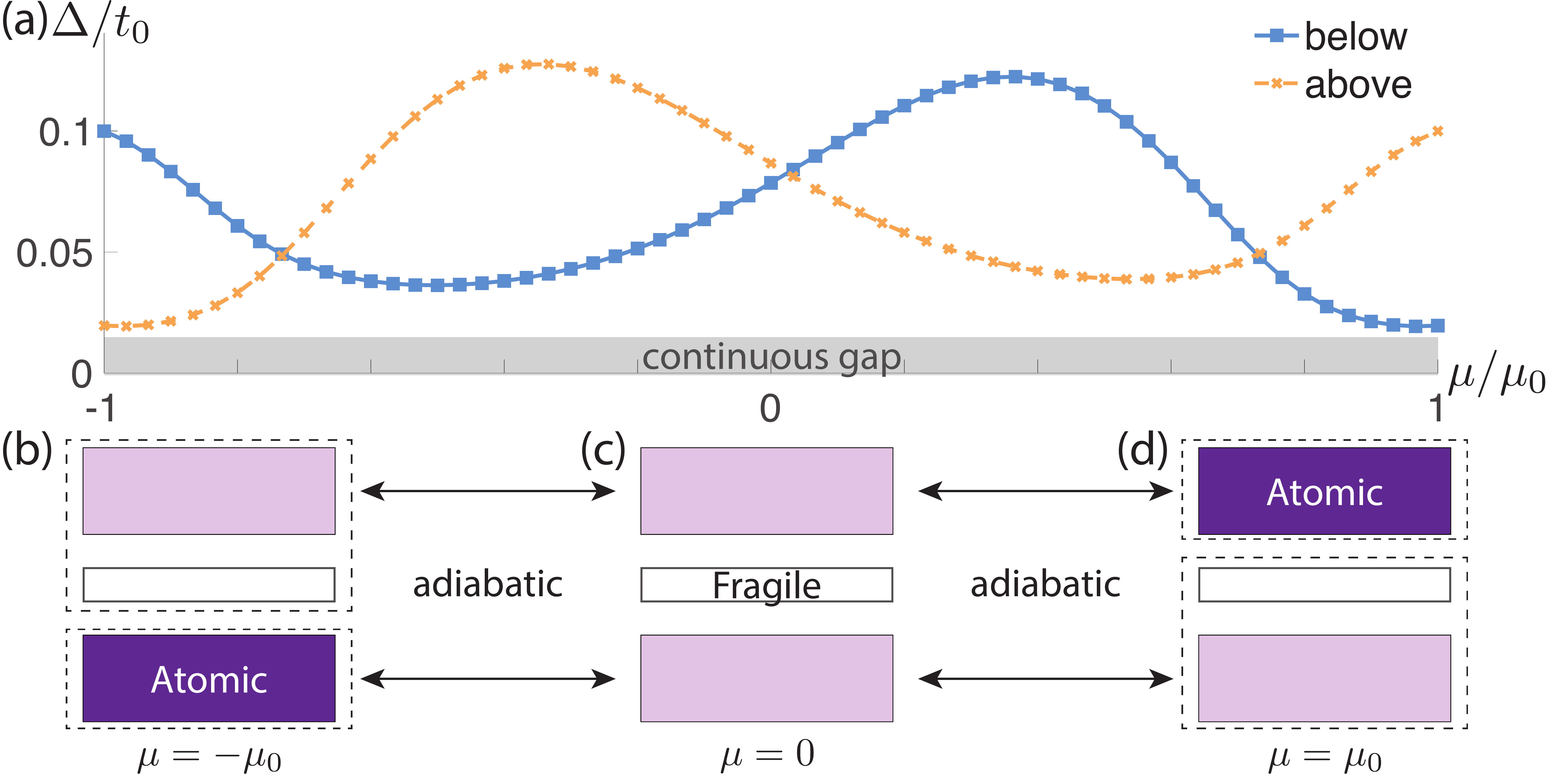}}
\caption{
Deformation to atomic limits. (a) The band gaps $\Delta$ below and above the two nearly flat bands stay open for all values of $\mu \in [-\mu_0,\mu_0]$ in Eq.\ \eqref{eq:Deform}.
More than $4\times 10^4$ momenta are sampled in the Brillouin zone in determining $\Delta$. We choose $\mu_0 = t_0 = 130$ meV.
(b--d) Schematic band diagrams at various limits of the deformation. (b) When $\mu = - \mu_0$, the lowest four bands arise solely from the $(\eta, p_\pm)$ orbitals and are therefore strictly atomic. Correspondingly, the upper six bands, altogether, are also in a strict atomic limit. (Dashed boxes indicate strictly atomic bands.)
The same is true for the case of $\mu = \mu_0$ in (d), but with the role of the lowest and highest bands exchanged. Since the band gaps are maintained throughout the entire deformation, we can infer that all the (light and dark) purple blocks correspond to trivial, atomic bands.
The nontrivial nearly flat bands at charge neutrality, therefore, must feature fragile topology.
\label{fig:Deform}
 }
\end{center}
\end{figure}

We have presented three tight-binding models, each containing two isolated bands with all the known band topology in TBG. At a glimpse, this might appear to follow the general phenomenology of topological bands, which can only arise in a tight-binding model when the topological invariants are neutralized by complementary bands possessing the ``opposite topology.''
Paradoxically, the complementary bands in our model are constructed using quasi-orbitals $h$ which correspond to an atomic insulator.
This indicates that the complementary bands in the ten-band model $\hat H(t_0,1)$ are all trivial, in that they can be smoothly deformed into explicit atomic insulators. 
More precisely, we will demonstrate this following a trick described in Ref.\ \onlinecite{Fragile}, which relies on a deformation Hamiltonian
\begin{equation}\begin{split}\label{eq:Deform}
\hat H'_{\mu} = &  \hat H \left ( f_\mu \, t_0 , f_\mu^2  \right) \\
& +\mu \sum_{\vec r, l=A,B, \alpha = \pm } \left(
\frac{1}{10}\hat \eta^{(l)\dagger}_{p_\alpha; \vec r }\hat \eta^{(l)}_{p_\alpha; \vec r }
-\hat h^{(l)\dagger}_{p_\alpha; \vec r }\hat h^{(l)}_{p_\alpha; \vec r }
\right),
\end{split}\end{equation}
where we choose the dimensionless function $f_\mu =   \cos(  \pi \mu / 2 \mu_0)$ such that $f_0 = 1$ and $f_{-\mu_0} = f_{\mu_0} = 0$, implying $\hat H'_{\mu=0}  = \hat H(t_0, 1)$. Note that the numerical factor of $1/10$ is {\it ad hoc} and is included simply to match the energy scales of the band gaps. Similarly, the precise form of $f_\mu$, as well as the appearance of $f_\mu^2$, have little physical meaning; these are just convenient choices that suffice for our purpose.
For $\mu = \mu_0>0$, the four highest bands coincide exactly with the atomic insulator arising from the full-filling of the $\eta_{p_{\pm}}$ orbitals, and the same is true for the  four lowest bands when $\mu = - \mu_0$. As shown in Fig.\ \ref{fig:Deform}a, the two band gaps in the spectrum never collapse for all $\mu \in [-\mu_0, \mu_0]$. This provides the needed adiabatic deformation to the explicit atomic limits (Fig.\ \ref{fig:Deform}b).

Curiously, as both the full tight-binding model as well as the complementary bands correspond to atomic insulators, the band topology of the two nearly flat bands conforms to the following equation:
 \begin{equation}\begin{split}\label{eq:fragile}
(\text{trivial}) = (\text{trivial}') \oplus (\text{nontrivial}),
\end{split}\end{equation}
where we say a set of bands is trivial if and only if they admit a full set of symmetric, localized Wannier functions, i.e., the full-filling of which leads to an insulator which can be smoothly deformed into a strict atomic limit \cite{Z2Wannier, SciAdv, NC, TopoChem}.
Eq.\ \eqref{eq:fragile} is the defining property of ``fragile topology'' \cite{Fragile,Cano2018,Slager2018}.
More concretely, we say the band topology of a set of gapped nontrivial bands is fragile if and only if one can append to the set another trivial set of bands such that, altogether, the augmented set is trivial; otherwise, we say the band topology is stable. As defined, stable and fragile topology are mutually exclusive concepts \cite{Fragile}.

Since the band topology of our ten-band model conforms to Eq.\ \eqref{eq:fragile},
our model also serves to prove that the identified form of band topology is fragile in nature.
For completeness, in Appendix \ref{app:Deform} we establish that the band topologies in our six-band and five-band models are also manifestly fragile.
This suggests that the interesting correlated behavior observed in TBG \cite{Cao2018a, Cao2018} could be related to interacting electrons occupying bands with an unconventional form of band topology.

Let us make two conceptual remarks before we close. First, we note that the fragile phenomenology of our models will persist as long as we retain $C_2 \mathcal T$ and lattice translations---the symmetries protecting the Dirac points. Therefore, upon the breaking of $M_y$ and $C_3$, our models provide an example of fragile topology not diagnosable using methods reliant on symmetry representations
\cite{NC,TopoChem}.
$C_2\mathcal T$-protected band topology has been studied in earlier works \cite{Fang2015, Alexandradinata2016, Ahn2018, Slager2018}. In particular, as a corollary from the fragile nature of our models, we remark that the 2D ``Stiefel-Whitney insulators'' proposed in Ref.\ \onlinecite{Ahn2018} can be atomic.

Second, in this work, we have provided three tight-binding models the properties of which are summarized in Table \ref{tab:TBSummary}. In particular, all three models have the property that the complementary bands are atomic in nature, as we demonstrate explicitly in Appendix \ref{app:Deform}.
Such models should be contrasted with the four-band model we introduced in Ref.\ \onlinecite{Jun2018}, as the complementary bands there have the same topological properties of the active bands in TBG. 
An interesting open question is whether or not the unconventional fragile nature of the band topology in TBG intertwines with the correlated physics. To answer this question, it is desirable to study models like those introduced in the present paper, for which the fragile nature of the bands is manifest. This then raises a conceptual question: what is the minimum number of bands required in constructing such models?
This question can be answered by combining analyses on symmetries (Appendix \ref{app:SymRep}) and band topology (Appendix \ref{app:Wilson}). The detailed argument on how to determine the minimum number of bands required can be found in Appendix \ref{app:Wilson}; here, we merely quote the result: at least five bands are required, and, therefore, the five-band model we introduced is minimal in this regard.

\section{Discussion}
In this work, we have critically examined the origin of the Wannier obstruction of the nearly flat bands in TBG, and identify it as fragile in nature \cite{Fragile}. The understanding of this Wannier obstruction is key to constructing tight-binding models which are important for future theoretical works on TBG. In contrast to a Wannier obstruction arising from stable topology, which necessarily requires bands with the opposite topological invariant to be supplied to resolve the obstruction,  here we have shown that the inclusion of trivial atomic bands is sufficient to resolve the fragile obstruction in TBG.
This is achieved by constructing tight-binding models that are both faithful and realistic, in that the additional bands not only resolve the Wannier obstruction, but also capture the energetics of nearby electronic bands.

Our constructions are based on the observation encapsulated in Eq.\ \eqref{eq:RepMatch} and its similarity with Eq.\ \eqref{eq:fragile}, where the latter is the defining phenomenology of fragile topology. However, we caution that such representation-matching equations are not unique (see Appendix \ref{app:SymRep} for more details). For instance, one can interchange $s \leftrightarrow p_z$ in Eq.\ \eqref{eq:RepMatch} and it still holds.
Alternatively, one might also swap some of the orbitals used through the equality $(\tau, p_z)  \oplus (\tau, p_{\pm}) \overset{{\rm rep.}}{=} (\kappa, s)$ (and similarly with $s\leftrightarrow p_z$), despite the fact that the two sides of this equation are not adiabatically deformable into one another \cite{Zak2001, TopoChem}. 
As a concrete example, we utilize a different representation-matching equation to construct a five-band model in Appendix \ref{app:5Band}. The smaller number of bands, however, comes at the cost of not faithfully capturing the representations of the higher-energy states, as we showed in the comparison in Table \ref{tab:TBSummary}.

The natural next step is to derive the dominant interaction terms in the problem by connecting our model to the microscopic degrees of freedom in TBG. Ideally, we would like to isolate the relevant bands in the continuum model and construct Wannier states  which  correspond to our tight-binding orbitals. We caution, however, that band crossings may occur at higher energies, so that isolating the relevant bands may require some judgment, although we do not expect this to affect the low-energy physics. More concretely, even if it is not possible to naturally identify a set of higher energy bands in TBG which are separated by band gaps and are atomic in nature, one can still disentangle effective set of bands with the desired properties from the rest of high-energy states \cite{VanderbiltRMP}. Importantly, these bands can be constructed in such a way that they differ from the actual energy eigenstates only at high energy, and therefore does not affect the low-energy property of the resulting model.

Another interesting future direction is to study how the unconventional nature of fragile topology might inform the physics of the interaction problem. Since the complementary trivial bands are fully filled, they correspond to an atomic insulator. In the limit where the band gap is much larger than the interaction strength, the problem should reduce to one involving certain local constraints on the Hilbert space. We leave these questions for future works.

{\it Note added}: In finalizing this manuscript, Ref.\ \onlinecite{Song2018} appeared, which proposes a four-band (per valley and per spin) tight-binding model similar to that described in Ref.\ \onlinecite{Jun2018}. They also suggest that the band topology in TBG is stable rather than fragile, based on the properties of Wilson loops  \cite{Fang2015, Alexandradinata2016,Ahn2018}.
As shown in Appendix \ref{app:Wilson}, the nearly flat bands in our tight-binding models have the same nontrivial Wilson loop invariant as identified in Ref.\ \onlinecite{Song2018}. However, we have explicitly demonstrated that their band topology is fragile. As stable and fragile topologies are mutually exclusive, this implies that the known Wannier obstructions in TBG cannot be stable.

\begin{acknowledgements}
We thank B.\ A.\ Bernevig, C.\ Fang, S.\ Fang, E.\ Kaxiras, R.-J, Slager, and M.\ Zaletel for helpful discussions.
TS is supported by a US
Department of Energy grant DE-SC0008739, and in
part by a Simons Investigator award from the Simons
Foundation. AV was supported by a Simons Investigator
award and by NSF-DMR 1411343. HCP was supported by a 
Pappalardo Fellowship at MIT and a Croucher Fellowship for
Postdoctoral research.
\end{acknowledgements}

\bibliography{TwBLGMott}

\begin{thebibliography}{68}%
\makeatletter
\providecommand \@ifxundefined [1]{%
 \@ifx{#1\undefined}
}%
\providecommand \@ifnum [1]{%
 \ifnum #1\expandafter \@firstoftwo
 \else \expandafter \@secondoftwo
 \fi
}%
\providecommand \@ifx [1]{%
 \ifx #1\expandafter \@firstoftwo
 \else \expandafter \@secondoftwo
 \fi
}%
\providecommand \natexlab [1]{#1}%
\providecommand \enquote  [1]{``#1''}%
\providecommand \bibnamefont  [1]{#1}%
\providecommand \bibfnamefont [1]{#1}%
\providecommand \citenamefont [1]{#1}%
\providecommand \href@noop [0]{\@secondoftwo}%
\providecommand \href [0]{\begingroup \@sanitize@url \@href}%
\providecommand \@href[1]{\@@startlink{#1}\@@href}%
\providecommand \@@href[1]{\endgroup#1\@@endlink}%
\providecommand \@sanitize@url [0]{\catcode `\\12\catcode `\$12\catcode
  `\&12\catcode `\#12\catcode `\^12\catcode `\_12\catcode `\%12\relax}%
\providecommand \@@startlink[1]{}%
\providecommand \@@endlink[0]{}%
\providecommand \url  [0]{\begingroup\@sanitize@url \@url }%
\providecommand \@url [1]{\endgroup\@href {#1}{\urlprefix }}%
\providecommand \urlprefix  [0]{URL }%
\providecommand \Eprint [0]{\href }%
\providecommand \doibase [0]{http://dx.doi.org/}%
\providecommand \selectlanguage [0]{\@gobble}%
\providecommand \bibinfo  [0]{\@secondoftwo}%
\providecommand \bibfield  [0]{\@secondoftwo}%
\providecommand \translation [1]{[#1]}%
\providecommand \BibitemOpen [0]{}%
\providecommand \bibitemStop [0]{}%
\providecommand \bibitemNoStop [0]{.\EOS\space}%
\providecommand \EOS [0]{\spacefactor3000\relax}%
\providecommand \BibitemShut  [1]{\csname bibitem#1\endcsname}%
\let\auto@bib@innerbib\@empty
\bibitem [{\citenamefont {Lee}\ \emph {et~al.}(2006)\citenamefont {Lee},
  \citenamefont {Nagaosa},\ and\ \citenamefont {Wen}}]{Lee2006}%
  \BibitemOpen
  \bibfield  {author} {\bibinfo {author} {\bibfnamefont {P.~A.}\ \bibnamefont
  {Lee}}, \bibinfo {author} {\bibfnamefont {N.}~\bibnamefont {Nagaosa}}, \ and\
  \bibinfo {author} {\bibfnamefont {X.-G.}\ \bibnamefont {Wen}},\ }\href@noop
  {} {\bibfield  {journal} {\bibinfo  {journal} {Rev. Mod. Phys.}\ }\textbf
  {\bibinfo {volume} {78}},\ \bibinfo {pages} {17} (\bibinfo {year}
  {2006})}\BibitemShut {NoStop}%
\bibitem [{\citenamefont {Cao}\ \emph {et~al.}(2018{\natexlab{a}})\citenamefont
  {Cao}, \citenamefont {Fatemi}, \citenamefont {Demir}, \citenamefont {Fang},
  \citenamefont {Tomarken}, \citenamefont {Luo}, \citenamefont
  {Sanchez-Yamagishi}, \citenamefont {Watanabe}, \citenamefont {Taniguchi},
  \citenamefont {Kaxiras}, \citenamefont {Ashoori},\ and\ \citenamefont
  {Jarillo-Herrero}}]{Cao2018}%
  \BibitemOpen
  \bibfield  {author} {\bibinfo {author} {\bibfnamefont {Y.}~\bibnamefont
  {Cao}}, \bibinfo {author} {\bibfnamefont {V.}~\bibnamefont {Fatemi}},
  \bibinfo {author} {\bibfnamefont {A.}~\bibnamefont {Demir}}, \bibinfo
  {author} {\bibfnamefont {S.}~\bibnamefont {Fang}}, \bibinfo {author}
  {\bibfnamefont {S.~L.}\ \bibnamefont {Tomarken}}, \bibinfo {author}
  {\bibfnamefont {J.~Y.}\ \bibnamefont {Luo}}, \bibinfo {author} {\bibfnamefont
  {J.~D.}\ \bibnamefont {Sanchez-Yamagishi}}, \bibinfo {author} {\bibfnamefont
  {K.}~\bibnamefont {Watanabe}}, \bibinfo {author} {\bibfnamefont
  {T.}~\bibnamefont {Taniguchi}}, \bibinfo {author} {\bibfnamefont
  {E.}~\bibnamefont {Kaxiras}}, \bibinfo {author} {\bibfnamefont {R.~C.}\
  \bibnamefont {Ashoori}}, \ and\ \bibinfo {author} {\bibfnamefont
  {P.}~\bibnamefont {Jarillo-Herrero}},\ }\href
  {http://dx.doi.org/10.1038/nature26154} {\bibfield  {journal} {\bibinfo
  {journal} {Nature}\ ,\ \bibinfo {pages} {80 }} (\bibinfo {year}
  {2018}{\natexlab{a}})}\BibitemShut {NoStop}%
\bibitem [{\citenamefont {Cao}\ \emph {et~al.}(2018{\natexlab{b}})\citenamefont
  {Cao}, \citenamefont {Fatemi}, \citenamefont {Fang}, \citenamefont
  {Watanabe}, \citenamefont {Taniguchi}, \citenamefont {Kaxiras},\ and\
  \citenamefont {Jarillo-Herrero}}]{Cao2018a}%
  \BibitemOpen
  \bibfield  {author} {\bibinfo {author} {\bibfnamefont {Y.}~\bibnamefont
  {Cao}}, \bibinfo {author} {\bibfnamefont {V.}~\bibnamefont {Fatemi}},
  \bibinfo {author} {\bibfnamefont {S.}~\bibnamefont {Fang}}, \bibinfo {author}
  {\bibfnamefont {K.}~\bibnamefont {Watanabe}}, \bibinfo {author}
  {\bibfnamefont {T.}~\bibnamefont {Taniguchi}}, \bibinfo {author}
  {\bibfnamefont {E.}~\bibnamefont {Kaxiras}}, \ and\ \bibinfo {author}
  {\bibfnamefont {P.}~\bibnamefont {Jarillo-Herrero}},\ }\href
  {http://dx.doi.org/10.1038/nature26160} {\bibfield  {journal} {\bibinfo
  {journal} {Nature}\ ,\ \bibinfo {pages} {43 }} (\bibinfo {year}
  {2018}{\natexlab{b}})}\BibitemShut {NoStop}%
\bibitem [{\citenamefont {Lopes~dos Santos}\ \emph {et~al.}(2007)\citenamefont
  {Lopes~dos Santos}, \citenamefont {Peres},\ and\ \citenamefont
  {Castro~Neto}}]{Neto2007}%
  \BibitemOpen
  \bibfield  {author} {\bibinfo {author} {\bibfnamefont {J.~M.~B.}\
  \bibnamefont {Lopes~dos Santos}}, \bibinfo {author} {\bibfnamefont
  {N.~M.~R.}\ \bibnamefont {Peres}}, \ and\ \bibinfo {author} {\bibfnamefont
  {A.~H.}\ \bibnamefont {Castro~Neto}},\ }\href {\doibase
  10.1103/PhysRevLett.99.256802} {\bibfield  {journal} {\bibinfo  {journal}
  {Phys. Rev. Lett.}\ }\textbf {\bibinfo {volume} {99}},\ \bibinfo {pages}
  {256802} (\bibinfo {year} {2007})}\BibitemShut {NoStop}%
\bibitem [{\citenamefont {Li}\ \emph {et~al.}(2009)\citenamefont {Li},
  \citenamefont {Luican}, \citenamefont {Lopes~dos Santos}, \citenamefont
  {Castro~Neto}, \citenamefont {Reina}, \citenamefont {Kong},\ and\
  \citenamefont {Andrei}}]{STMNatPhy}%
  \BibitemOpen
  \bibfield  {author} {\bibinfo {author} {\bibfnamefont {G.}~\bibnamefont
  {Li}}, \bibinfo {author} {\bibfnamefont {A.}~\bibnamefont {Luican}}, \bibinfo
  {author} {\bibfnamefont {J.~M.~B.}\ \bibnamefont {Lopes~dos Santos}},
  \bibinfo {author} {\bibfnamefont {A.~H.}\ \bibnamefont {Castro~Neto}},
  \bibinfo {author} {\bibfnamefont {A.}~\bibnamefont {Reina}}, \bibinfo
  {author} {\bibfnamefont {J.}~\bibnamefont {Kong}}, \ and\ \bibinfo {author}
  {\bibfnamefont {E.~Y.}\ \bibnamefont {Andrei}},\ }\href
  {http://dx.doi.org/10.1038/nphys1463} {\bibfield  {journal} {\bibinfo
  {journal} {Nature Physics}\ }\textbf {\bibinfo {volume} {6}},\ \bibinfo
  {pages} {109 } (\bibinfo {year} {2009})}\BibitemShut {NoStop}%
\bibitem [{\citenamefont {Trambly~de Laissardi{\`e}re}\ \emph
  {et~al.}(2010)\citenamefont {Trambly~de Laissardi{\`e}re}, \citenamefont
  {Mayou},\ and\ \citenamefont {Magaud}}]{Magaud2010}%
  \BibitemOpen
  \bibfield  {author} {\bibinfo {author} {\bibfnamefont {G.}~\bibnamefont
  {Trambly~de Laissardi{\`e}re}}, \bibinfo {author} {\bibfnamefont
  {D.}~\bibnamefont {Mayou}}, \ and\ \bibinfo {author} {\bibfnamefont
  {L.}~\bibnamefont {Magaud}},\ }\bibfield  {booktitle} {\emph {\bibinfo
  {booktitle} {Nano Letters}},\ }\href {\doibase 10.1021/nl902948m} {\bibfield
  {journal} {\bibinfo  {journal} {Nano Letters}\ }\textbf {\bibinfo {volume}
  {10}},\ \bibinfo {pages} {804} (\bibinfo {year} {2010})}\BibitemShut
  {NoStop}%
\bibitem [{\citenamefont {Mele}(2010)}]{Mele2010}%
  \BibitemOpen
  \bibfield  {author} {\bibinfo {author} {\bibfnamefont {E.~J.}\ \bibnamefont
  {Mele}},\ }\href {\doibase 10.1103/PhysRevB.81.161405} {\bibfield  {journal}
  {\bibinfo  {journal} {Phys. Rev. B}\ }\textbf {\bibinfo {volume} {81}},\
  \bibinfo {pages} {161405} (\bibinfo {year} {2010})}\BibitemShut {NoStop}%
\bibitem [{\citenamefont {Shallcross}\ \emph {et~al.}(2010)\citenamefont
  {Shallcross}, \citenamefont {Sharma}, \citenamefont {Kandelaki},\ and\
  \citenamefont {Pankratov}}]{Shallcross2010}%
  \BibitemOpen
  \bibfield  {author} {\bibinfo {author} {\bibfnamefont {S.}~\bibnamefont
  {Shallcross}}, \bibinfo {author} {\bibfnamefont {S.}~\bibnamefont {Sharma}},
  \bibinfo {author} {\bibfnamefont {E.}~\bibnamefont {Kandelaki}}, \ and\
  \bibinfo {author} {\bibfnamefont {O.~A.}\ \bibnamefont {Pankratov}},\ }\href
  {\doibase 10.1103/PhysRevB.81.165105} {\bibfield  {journal} {\bibinfo
  {journal} {Phys. Rev. B}\ }\textbf {\bibinfo {volume} {81}},\ \bibinfo
  {pages} {165105} (\bibinfo {year} {2010})}\BibitemShut {NoStop}%
\bibitem [{\citenamefont {Bistritzer}\ and\ \citenamefont
  {MacDonald}(2010)}]{Bistritzer2010}%
  \BibitemOpen
  \bibfield  {author} {\bibinfo {author} {\bibfnamefont {R.}~\bibnamefont
  {Bistritzer}}\ and\ \bibinfo {author} {\bibfnamefont {A.~H.}\ \bibnamefont
  {MacDonald}},\ }\href {\doibase 10.1103/PhysRevB.81.245412} {\bibfield
  {journal} {\bibinfo  {journal} {Phys. Rev. B}\ }\textbf {\bibinfo {volume}
  {81}},\ \bibinfo {pages} {245412} (\bibinfo {year} {2010})}\BibitemShut
  {NoStop}%
\bibitem [{\citenamefont {Su\'arez~Morell}\ \emph {et~al.}(2010)\citenamefont
  {Su\'arez~Morell}, \citenamefont {Correa}, \citenamefont {Vargas},
  \citenamefont {Pacheco},\ and\ \citenamefont {Barticevic}}]{Morell2010}%
  \BibitemOpen
  \bibfield  {author} {\bibinfo {author} {\bibfnamefont {E.}~\bibnamefont
  {Su\'arez~Morell}}, \bibinfo {author} {\bibfnamefont {J.~D.}\ \bibnamefont
  {Correa}}, \bibinfo {author} {\bibfnamefont {P.}~\bibnamefont {Vargas}},
  \bibinfo {author} {\bibfnamefont {M.}~\bibnamefont {Pacheco}}, \ and\
  \bibinfo {author} {\bibfnamefont {Z.}~\bibnamefont {Barticevic}},\ }\href
  {\doibase 10.1103/PhysRevB.82.121407} {\bibfield  {journal} {\bibinfo
  {journal} {Phys. Rev. B}\ }\textbf {\bibinfo {volume} {82}},\ \bibinfo
  {pages} {121407} (\bibinfo {year} {2010})}\BibitemShut {NoStop}%
\bibitem [{\citenamefont {Luican}\ \emph {et~al.}(2011)\citenamefont {Luican},
  \citenamefont {Li}, \citenamefont {Reina}, \citenamefont {Kong},
  \citenamefont {Nair}, \citenamefont {Novoselov}, \citenamefont {Geim},\ and\
  \citenamefont {Andrei}}]{STMPRL}%
  \BibitemOpen
  \bibfield  {author} {\bibinfo {author} {\bibfnamefont {A.}~\bibnamefont
  {Luican}}, \bibinfo {author} {\bibfnamefont {G.}~\bibnamefont {Li}}, \bibinfo
  {author} {\bibfnamefont {A.}~\bibnamefont {Reina}}, \bibinfo {author}
  {\bibfnamefont {J.}~\bibnamefont {Kong}}, \bibinfo {author} {\bibfnamefont
  {R.~R.}\ \bibnamefont {Nair}}, \bibinfo {author} {\bibfnamefont {K.~S.}\
  \bibnamefont {Novoselov}}, \bibinfo {author} {\bibfnamefont {A.~K.}\
  \bibnamefont {Geim}}, \ and\ \bibinfo {author} {\bibfnamefont {E.~Y.}\
  \bibnamefont {Andrei}},\ }\href {\doibase 10.1103/PhysRevLett.106.126802}
  {\bibfield  {journal} {\bibinfo  {journal} {Phys. Rev. Lett.}\ }\textbf
  {\bibinfo {volume} {106}},\ \bibinfo {pages} {126802} (\bibinfo {year}
  {2011})}\BibitemShut {NoStop}%
\bibitem [{\citenamefont {Bistritzer}\ and\ \citenamefont
  {MacDonald}(2011)}]{Bistritzer2011}%
  \BibitemOpen
  \bibfield  {author} {\bibinfo {author} {\bibfnamefont {R.}~\bibnamefont
  {Bistritzer}}\ and\ \bibinfo {author} {\bibfnamefont {A.~H.}\ \bibnamefont
  {MacDonald}},\ }\href {\doibase 10.1073/pnas.1108174108} {\bibfield
  {journal} {\bibinfo  {journal} {Proceedings of the National Academy of
  Sciences}\ }\textbf {\bibinfo {volume} {108}},\ \bibinfo {pages} {12233}
  (\bibinfo {year} {2011})},\ \Eprint
  {http://arxiv.org/abs/http://www.pnas.org/content/108/30/12233.full.pdf}
  {http://www.pnas.org/content/108/30/12233.full.pdf} \BibitemShut {NoStop}%
\bibitem [{\citenamefont {de~Gail}\ \emph {et~al.}(2011)\citenamefont
  {de~Gail}, \citenamefont {Goerbig}, \citenamefont {Guinea}, \citenamefont
  {Montambaux},\ and\ \citenamefont {Castro~Neto}}]{CastroNeto2011}%
  \BibitemOpen
  \bibfield  {author} {\bibinfo {author} {\bibfnamefont {R.}~\bibnamefont
  {de~Gail}}, \bibinfo {author} {\bibfnamefont {M.~O.}\ \bibnamefont
  {Goerbig}}, \bibinfo {author} {\bibfnamefont {F.}~\bibnamefont {Guinea}},
  \bibinfo {author} {\bibfnamefont {G.}~\bibnamefont {Montambaux}}, \ and\
  \bibinfo {author} {\bibfnamefont {A.~H.}\ \bibnamefont {Castro~Neto}},\
  }\href {\doibase 10.1103/PhysRevB.84.045436} {\bibfield  {journal} {\bibinfo
  {journal} {Phys. Rev. B}\ }\textbf {\bibinfo {volume} {84}},\ \bibinfo
  {pages} {045436} (\bibinfo {year} {2011})}\BibitemShut {NoStop}%
\bibitem [{\citenamefont {Lopes~dos Santos}\ \emph {et~al.}(2012)\citenamefont
  {Lopes~dos Santos}, \citenamefont {Peres},\ and\ \citenamefont
  {Castro~Neto}}]{Castro-Neto2012}%
  \BibitemOpen
  \bibfield  {author} {\bibinfo {author} {\bibfnamefont {J.~M.~B.}\
  \bibnamefont {Lopes~dos Santos}}, \bibinfo {author} {\bibfnamefont
  {N.~M.~R.}\ \bibnamefont {Peres}}, \ and\ \bibinfo {author} {\bibfnamefont
  {A.~H.}\ \bibnamefont {Castro~Neto}},\ }\href {\doibase
  10.1103/PhysRevB.86.155449} {\bibfield  {journal} {\bibinfo  {journal} {Phys.
  Rev. B}\ }\textbf {\bibinfo {volume} {86}},\ \bibinfo {pages} {155449}
  (\bibinfo {year} {2012})}\BibitemShut {NoStop}%
\bibitem [{\citenamefont {He}\ \emph {et~al.}(2013)\citenamefont {He},
  \citenamefont {Chu},\ and\ \citenamefont {He}}]{He2013}%
  \BibitemOpen
  \bibfield  {author} {\bibinfo {author} {\bibfnamefont {W.-Y.}\ \bibnamefont
  {He}}, \bibinfo {author} {\bibfnamefont {Z.-D.}\ \bibnamefont {Chu}}, \ and\
  \bibinfo {author} {\bibfnamefont {L.}~\bibnamefont {He}},\ }\href {\doibase
  10.1103/PhysRevLett.111.066803} {\bibfield  {journal} {\bibinfo  {journal}
  {Phys. Rev. Lett.}\ }\textbf {\bibinfo {volume} {111}},\ \bibinfo {pages}
  {066803} (\bibinfo {year} {2013})}\BibitemShut {NoStop}%
\bibitem [{\citenamefont {Uchida}\ \emph {et~al.}(2014)\citenamefont {Uchida},
  \citenamefont {Furuya}, \citenamefont {Iwata},\ and\ \citenamefont
  {Oshiyama}}]{Uchida2014}%
  \BibitemOpen
  \bibfield  {author} {\bibinfo {author} {\bibfnamefont {K.}~\bibnamefont
  {Uchida}}, \bibinfo {author} {\bibfnamefont {S.}~\bibnamefont {Furuya}},
  \bibinfo {author} {\bibfnamefont {J.-I.}\ \bibnamefont {Iwata}}, \ and\
  \bibinfo {author} {\bibfnamefont {A.}~\bibnamefont {Oshiyama}},\ }\href
  {\doibase 10.1103/PhysRevB.90.155451} {\bibfield  {journal} {\bibinfo
  {journal} {Phys. Rev. B}\ }\textbf {\bibinfo {volume} {90}},\ \bibinfo
  {pages} {155451} (\bibinfo {year} {2014})}\BibitemShut {NoStop}%
\bibitem [{\citenamefont {Yan}\ \emph {et~al.}(2012)\citenamefont {Yan},
  \citenamefont {Liu}, \citenamefont {Dou}, \citenamefont {Meng}, \citenamefont
  {Feng}, \citenamefont {Chu}, \citenamefont {Zhang}, \citenamefont {Liu},
  \citenamefont {Nie},\ and\ \citenamefont {He}}]{Yan2012}%
  \BibitemOpen
  \bibfield  {author} {\bibinfo {author} {\bibfnamefont {W.}~\bibnamefont
  {Yan}}, \bibinfo {author} {\bibfnamefont {M.}~\bibnamefont {Liu}}, \bibinfo
  {author} {\bibfnamefont {R.-F.}\ \bibnamefont {Dou}}, \bibinfo {author}
  {\bibfnamefont {L.}~\bibnamefont {Meng}}, \bibinfo {author} {\bibfnamefont
  {L.}~\bibnamefont {Feng}}, \bibinfo {author} {\bibfnamefont {Z.-D.}\
  \bibnamefont {Chu}}, \bibinfo {author} {\bibfnamefont {Y.}~\bibnamefont
  {Zhang}}, \bibinfo {author} {\bibfnamefont {Z.}~\bibnamefont {Liu}}, \bibinfo
  {author} {\bibfnamefont {J.-C.}\ \bibnamefont {Nie}}, \ and\ \bibinfo
  {author} {\bibfnamefont {L.}~\bibnamefont {He}},\ }\href {\doibase
  10.1103/PhysRevLett.109.126801} {\bibfield  {journal} {\bibinfo  {journal}
  {Phys. Rev. Lett.}\ }\textbf {\bibinfo {volume} {109}},\ \bibinfo {pages}
  {126801} (\bibinfo {year} {2012})}\BibitemShut {NoStop}%
\bibitem [{\citenamefont {Jung}\ \emph {et~al.}(2014)\citenamefont {Jung},
  \citenamefont {Raoux}, \citenamefont {Qiao},\ and\ \citenamefont
  {MacDonald}}]{Jung2014}%
  \BibitemOpen
  \bibfield  {author} {\bibinfo {author} {\bibfnamefont {J.}~\bibnamefont
  {Jung}}, \bibinfo {author} {\bibfnamefont {A.}~\bibnamefont {Raoux}},
  \bibinfo {author} {\bibfnamefont {Z.}~\bibnamefont {Qiao}}, \ and\ \bibinfo
  {author} {\bibfnamefont {A.~H.}\ \bibnamefont {MacDonald}},\ }\href {\doibase
  10.1103/PhysRevB.89.205414} {\bibfield  {journal} {\bibinfo  {journal} {Phys.
  Rev. B}\ }\textbf {\bibinfo {volume} {89}},\ \bibinfo {pages} {205414}
  (\bibinfo {year} {2014})}\BibitemShut {NoStop}%
\bibitem [{\citenamefont {Sboychakov}\ \emph {et~al.}(2015)\citenamefont
  {Sboychakov}, \citenamefont {Rakhmanov}, \citenamefont {Rozhkov},\ and\
  \citenamefont {Nori}}]{Sboychakov2015}%
  \BibitemOpen
  \bibfield  {author} {\bibinfo {author} {\bibfnamefont {A.~O.}\ \bibnamefont
  {Sboychakov}}, \bibinfo {author} {\bibfnamefont {A.~L.}\ \bibnamefont
  {Rakhmanov}}, \bibinfo {author} {\bibfnamefont {A.~V.}\ \bibnamefont
  {Rozhkov}}, \ and\ \bibinfo {author} {\bibfnamefont {F.}~\bibnamefont
  {Nori}},\ }\href {\doibase 10.1103/PhysRevB.92.075402} {\bibfield  {journal}
  {\bibinfo  {journal} {Phys. Rev. B}\ }\textbf {\bibinfo {volume} {92}},\
  \bibinfo {pages} {075402} (\bibinfo {year} {2015})}\BibitemShut {NoStop}%
\bibitem [{\citenamefont {Wong}\ \emph {et~al.}(2015)\citenamefont {Wong},
  \citenamefont {Wang}, \citenamefont {Jung}, \citenamefont {Pezzini},
  \citenamefont {DaSilva}, \citenamefont {Tsai}, \citenamefont {Jung},
  \citenamefont {Khajeh}, \citenamefont {Kim}, \citenamefont {Lee},
  \citenamefont {Kahn}, \citenamefont {Tollabimazraehno}, \citenamefont
  {Rasool}, \citenamefont {Watanabe}, \citenamefont {Taniguchi}, \citenamefont
  {Zettl}, \citenamefont {Adam}, \citenamefont {MacDonald},\ and\ \citenamefont
  {Crommie}}]{Crommie2015}%
  \BibitemOpen
  \bibfield  {author} {\bibinfo {author} {\bibfnamefont {D.}~\bibnamefont
  {Wong}}, \bibinfo {author} {\bibfnamefont {Y.}~\bibnamefont {Wang}}, \bibinfo
  {author} {\bibfnamefont {J.}~\bibnamefont {Jung}}, \bibinfo {author}
  {\bibfnamefont {S.}~\bibnamefont {Pezzini}}, \bibinfo {author} {\bibfnamefont
  {A.~M.}\ \bibnamefont {DaSilva}}, \bibinfo {author} {\bibfnamefont {H.-Z.}\
  \bibnamefont {Tsai}}, \bibinfo {author} {\bibfnamefont {H.~S.}\ \bibnamefont
  {Jung}}, \bibinfo {author} {\bibfnamefont {R.}~\bibnamefont {Khajeh}},
  \bibinfo {author} {\bibfnamefont {Y.}~\bibnamefont {Kim}}, \bibinfo {author}
  {\bibfnamefont {J.}~\bibnamefont {Lee}}, \bibinfo {author} {\bibfnamefont
  {S.}~\bibnamefont {Kahn}}, \bibinfo {author} {\bibfnamefont {S.}~\bibnamefont
  {Tollabimazraehno}}, \bibinfo {author} {\bibfnamefont {H.}~\bibnamefont
  {Rasool}}, \bibinfo {author} {\bibfnamefont {K.}~\bibnamefont {Watanabe}},
  \bibinfo {author} {\bibfnamefont {T.}~\bibnamefont {Taniguchi}}, \bibinfo
  {author} {\bibfnamefont {A.}~\bibnamefont {Zettl}}, \bibinfo {author}
  {\bibfnamefont {S.}~\bibnamefont {Adam}}, \bibinfo {author} {\bibfnamefont
  {A.~H.}\ \bibnamefont {MacDonald}}, \ and\ \bibinfo {author} {\bibfnamefont
  {M.~F.}\ \bibnamefont {Crommie}},\ }\href {\doibase
  10.1103/PhysRevB.92.155409} {\bibfield  {journal} {\bibinfo  {journal} {Phys.
  Rev. B}\ }\textbf {\bibinfo {volume} {92}},\ \bibinfo {pages} {155409}
  (\bibinfo {year} {2015})}\BibitemShut {NoStop}%
\bibitem [{\citenamefont {Fang}\ and\ \citenamefont
  {Kaxiras}(2016)}]{ShiangPRB}%
  \BibitemOpen
  \bibfield  {author} {\bibinfo {author} {\bibfnamefont {S.}~\bibnamefont
  {Fang}}\ and\ \bibinfo {author} {\bibfnamefont {E.}~\bibnamefont {Kaxiras}},\
  }\href {\doibase 10.1103/PhysRevB.93.235153} {\bibfield  {journal} {\bibinfo
  {journal} {Phys. Rev. B}\ }\textbf {\bibinfo {volume} {93}},\ \bibinfo
  {pages} {235153} (\bibinfo {year} {2016})}\BibitemShut {NoStop}%
\bibitem [{\citenamefont {Cao}\ \emph {et~al.}(2016)\citenamefont {Cao},
  \citenamefont {Luo}, \citenamefont {Fatemi}, \citenamefont {Fang},
  \citenamefont {Sanchez-Yamagishi}, \citenamefont {Watanabe}, \citenamefont
  {Taniguchi}, \citenamefont {Kaxiras},\ and\ \citenamefont
  {Jarillo-Herrero}}]{PabloPRL}%
  \BibitemOpen
  \bibfield  {author} {\bibinfo {author} {\bibfnamefont {Y.}~\bibnamefont
  {Cao}}, \bibinfo {author} {\bibfnamefont {J.~Y.}\ \bibnamefont {Luo}},
  \bibinfo {author} {\bibfnamefont {V.}~\bibnamefont {Fatemi}}, \bibinfo
  {author} {\bibfnamefont {S.}~\bibnamefont {Fang}}, \bibinfo {author}
  {\bibfnamefont {J.~D.}\ \bibnamefont {Sanchez-Yamagishi}}, \bibinfo {author}
  {\bibfnamefont {K.}~\bibnamefont {Watanabe}}, \bibinfo {author}
  {\bibfnamefont {T.}~\bibnamefont {Taniguchi}}, \bibinfo {author}
  {\bibfnamefont {E.}~\bibnamefont {Kaxiras}}, \ and\ \bibinfo {author}
  {\bibfnamefont {P.}~\bibnamefont {Jarillo-Herrero}},\ }\href {\doibase
  10.1103/PhysRevLett.117.116804} {\bibfield  {journal} {\bibinfo  {journal}
  {Phys. Rev. Lett.}\ }\textbf {\bibinfo {volume} {117}},\ \bibinfo {pages}
  {116804} (\bibinfo {year} {2016})}\BibitemShut {NoStop}%
\bibitem [{\citenamefont {Kim}\ \emph {et~al.}(2017)\citenamefont {Kim},
  \citenamefont {DaSilva}, \citenamefont {Huang}, \citenamefont {Fallahazad},
  \citenamefont {Larentis}, \citenamefont {Taniguchi}, \citenamefont
  {Watanabe}, \citenamefont {LeRoy}, \citenamefont {MacDonald},\ and\
  \citenamefont {Tutuc}}]{Kim2017}%
  \BibitemOpen
  \bibfield  {author} {\bibinfo {author} {\bibfnamefont {K.}~\bibnamefont
  {Kim}}, \bibinfo {author} {\bibfnamefont {A.}~\bibnamefont {DaSilva}},
  \bibinfo {author} {\bibfnamefont {S.}~\bibnamefont {Huang}}, \bibinfo
  {author} {\bibfnamefont {B.}~\bibnamefont {Fallahazad}}, \bibinfo {author}
  {\bibfnamefont {S.}~\bibnamefont {Larentis}}, \bibinfo {author}
  {\bibfnamefont {T.}~\bibnamefont {Taniguchi}}, \bibinfo {author}
  {\bibfnamefont {K.}~\bibnamefont {Watanabe}}, \bibinfo {author}
  {\bibfnamefont {B.~J.}\ \bibnamefont {LeRoy}}, \bibinfo {author}
  {\bibfnamefont {A.~H.}\ \bibnamefont {MacDonald}}, \ and\ \bibinfo {author}
  {\bibfnamefont {E.}~\bibnamefont {Tutuc}},\ }\href
  {http://www.pnas.org/content/114/13/3364.abstract} {\bibfield  {journal}
  {\bibinfo  {journal} {Proceedings of the National Academy of Sciences}\
  }\textbf {\bibinfo {volume} {114}},\ \bibinfo {pages} {3364} (\bibinfo {year}
  {2017})}\BibitemShut {NoStop}%
\bibitem [{\citenamefont {Nam}\ and\ \citenamefont {Koshino}(2017)}]{Nam2017}%
  \BibitemOpen
  \bibfield  {author} {\bibinfo {author} {\bibfnamefont {N.~N.~T.}\
  \bibnamefont {Nam}}\ and\ \bibinfo {author} {\bibfnamefont {M.}~\bibnamefont
  {Koshino}},\ }\href {\doibase 10.1103/PhysRevB.96.075311} {\bibfield
  {journal} {\bibinfo  {journal} {Phys. Rev. B}\ }\textbf {\bibinfo {volume}
  {96}},\ \bibinfo {pages} {075311} (\bibinfo {year} {2017})}\BibitemShut
  {NoStop}%
\bibitem [{\citenamefont {Huang}\ \emph {et~al.}(2018)\citenamefont {Huang},
  \citenamefont {Kim}, \citenamefont {Efimkin}, \citenamefont {Lovorn},
  \citenamefont {Taniguchi}, \citenamefont {Watanabe}, \citenamefont
  {MacDonald}, \citenamefont {Tutuc},\ and\ \citenamefont {LeRoy}}]{LeRoy2018}%
  \BibitemOpen
  \bibfield  {author} {\bibinfo {author} {\bibfnamefont {S.}~\bibnamefont
  {Huang}}, \bibinfo {author} {\bibfnamefont {K.}~\bibnamefont {Kim}}, \bibinfo
  {author} {\bibfnamefont {D.~K.}\ \bibnamefont {Efimkin}}, \bibinfo {author}
  {\bibfnamefont {T.}~\bibnamefont {Lovorn}}, \bibinfo {author} {\bibfnamefont
  {T.}~\bibnamefont {Taniguchi}}, \bibinfo {author} {\bibfnamefont
  {K.}~\bibnamefont {Watanabe}}, \bibinfo {author} {\bibfnamefont {A.~H.}\
  \bibnamefont {MacDonald}}, \bibinfo {author} {\bibfnamefont {E.}~\bibnamefont
  {Tutuc}}, \ and\ \bibinfo {author} {\bibfnamefont {B.~J.}\ \bibnamefont
  {LeRoy}},\ }\href {\doibase 10.1103/PhysRevLett.121.037702} {\bibfield
  {journal} {\bibinfo  {journal} {Phys. Rev. Lett.}\ }\textbf {\bibinfo
  {volume} {121}},\ \bibinfo {pages} {037702} (\bibinfo {year}
  {2018})}\BibitemShut {NoStop}%
\bibitem [{\citenamefont {Rickhaus}\ \emph {et~al.}(2018)\citenamefont
  {Rickhaus}, \citenamefont {Wallbank}, \citenamefont {Slizovskiy},
  \citenamefont {Pisoni}, \citenamefont {Overweg}, \citenamefont {Lee},
  \citenamefont {Eich}, \citenamefont {Liu}, \citenamefont {Watanabe},
  \citenamefont {Taniguchi}, \citenamefont {Ihn},\ and\ \citenamefont
  {Ensslin}}]{Ensslin2018}%
  \BibitemOpen
  \bibfield  {author} {\bibinfo {author} {\bibfnamefont {P.}~\bibnamefont
  {Rickhaus}}, \bibinfo {author} {\bibfnamefont {J.}~\bibnamefont {Wallbank}},
  \bibinfo {author} {\bibfnamefont {S.}~\bibnamefont {Slizovskiy}}, \bibinfo
  {author} {\bibfnamefont {R.}~\bibnamefont {Pisoni}}, \bibinfo {author}
  {\bibfnamefont {H.}~\bibnamefont {Overweg}}, \bibinfo {author} {\bibfnamefont
  {Y.}~\bibnamefont {Lee}}, \bibinfo {author} {\bibfnamefont {M.}~\bibnamefont
  {Eich}}, \bibinfo {author} {\bibfnamefont {M.-H.}\ \bibnamefont {Liu}},
  \bibinfo {author} {\bibfnamefont {K.}~\bibnamefont {Watanabe}}, \bibinfo
  {author} {\bibfnamefont {T.}~\bibnamefont {Taniguchi}}, \bibinfo {author}
  {\bibfnamefont {T.}~\bibnamefont {Ihn}}, \ and\ \bibinfo {author}
  {\bibfnamefont {K.}~\bibnamefont {Ensslin}},\ }\href {\doibase
  10.1021/acs.nanolett.8b02387} {\bibfield  {journal} {\bibinfo  {journal}
  {Nano Letters}\ }\textbf {\bibinfo {volume} {18}},\ \bibinfo {pages} {6725}
  (\bibinfo {year} {2018})},\ \bibinfo {note} {pMID: 30336041},\ \Eprint
  {http://arxiv.org/abs/https://doi.org/10.1021/acs.nanolett.8b02387}
  {https://doi.org/10.1021/acs.nanolett.8b02387} \BibitemShut {NoStop}%
\bibitem [{\citenamefont {Ramires}\ and\ \citenamefont {Lado}(2018)}]{Ramires}%
  \BibitemOpen
  \bibfield  {author} {\bibinfo {author} {\bibfnamefont {A.}~\bibnamefont
  {Ramires}}\ and\ \bibinfo {author} {\bibfnamefont {J.~L.}\ \bibnamefont
  {Lado}},\ }\href {\doibase 10.1103/PhysRevLett.121.146801} {\bibfield
  {journal} {\bibinfo  {journal} {Phys. Rev. Lett.}\ }\textbf {\bibinfo
  {volume} {121}},\ \bibinfo {pages} {146801} (\bibinfo {year}
  {2018})}\BibitemShut {NoStop}%
\bibitem [{\citenamefont {Po}\ \emph {et~al.}(2018{\natexlab{a}})\citenamefont
  {Po}, \citenamefont {Zou}, \citenamefont {Vishwanath},\ and\ \citenamefont
  {Senthil}}]{Mar2018}%
  \BibitemOpen
  \bibfield  {author} {\bibinfo {author} {\bibfnamefont {H.~C.}\ \bibnamefont
  {Po}}, \bibinfo {author} {\bibfnamefont {L.}~\bibnamefont {Zou}}, \bibinfo
  {author} {\bibfnamefont {A.}~\bibnamefont {Vishwanath}}, \ and\ \bibinfo
  {author} {\bibfnamefont {T.}~\bibnamefont {Senthil}},\ }\href {\doibase
  10.1103/PhysRevX.8.031089} {\bibfield  {journal} {\bibinfo  {journal} {Phys.
  Rev. X}\ }\textbf {\bibinfo {volume} {8}},\ \bibinfo {pages} {031089}
  (\bibinfo {year} {2018}{\natexlab{a}})}\BibitemShut {NoStop}%
\bibitem [{\citenamefont {Zou}\ \emph {et~al.}(2018)\citenamefont {Zou},
  \citenamefont {Po}, \citenamefont {Vishwanath},\ and\ \citenamefont
  {Senthil}}]{Jun2018}%
  \BibitemOpen
  \bibfield  {author} {\bibinfo {author} {\bibfnamefont {L.}~\bibnamefont
  {Zou}}, \bibinfo {author} {\bibfnamefont {H.~C.}\ \bibnamefont {Po}},
  \bibinfo {author} {\bibfnamefont {A.}~\bibnamefont {Vishwanath}}, \ and\
  \bibinfo {author} {\bibfnamefont {T.}~\bibnamefont {Senthil}},\ }\href
  {\doibase 10.1103/PhysRevB.98.085435} {\bibfield  {journal} {\bibinfo
  {journal} {Phys. Rev. B}\ }\textbf {\bibinfo {volume} {98}},\ \bibinfo
  {pages} {085435} (\bibinfo {year} {2018})}\BibitemShut {NoStop}%
\bibitem [{\citenamefont {Goerbig}\ and\ \citenamefont
  {Montambaux}(2017)}]{Goerbig2017}%
  \BibitemOpen
  \bibfield  {author} {\bibinfo {author} {\bibfnamefont {M.}~\bibnamefont
  {Goerbig}}\ and\ \bibinfo {author} {\bibfnamefont {G.}~\bibnamefont
  {Montambaux}},\ }\enquote {\bibinfo {title} {Dirac fermions in condensed
  matter and beyond},}\ in\ \href {\doibase 10.1007/978-3-319-32536-1_2} {\emph
  {\bibinfo {booktitle} {Dirac Matter}}},\ \bibinfo {editor} {edited by\
  \bibinfo {editor} {\bibfnamefont {B.}~\bibnamefont {Duplantier}}, \bibinfo
  {editor} {\bibfnamefont {V.}~\bibnamefont {Rivasseau}}, \ and\ \bibinfo
  {editor} {\bibfnamefont {J.-N.}\ \bibnamefont {Fuchs}}}\ (\bibinfo
  {publisher} {Springer International Publishing},\ \bibinfo {address} {Cham},\
  \bibinfo {year} {2017})\ pp.\ \bibinfo {pages} {25--53}\BibitemShut {NoStop}%
\bibitem [{\citenamefont {Soluyanov}\ and\ \citenamefont
  {Vanderbilt}(2011)}]{Z2Wannier}%
  \BibitemOpen
  \bibfield  {author} {\bibinfo {author} {\bibfnamefont {A.~A.}\ \bibnamefont
  {Soluyanov}}\ and\ \bibinfo {author} {\bibfnamefont {D.}~\bibnamefont
  {Vanderbilt}},\ }\href {\doibase 10.1103/PhysRevB.83.035108} {\bibfield
  {journal} {\bibinfo  {journal} {Phys. Rev. B}\ }\textbf {\bibinfo {volume}
  {83}},\ \bibinfo {pages} {035108} (\bibinfo {year} {2011})}\BibitemShut
  {NoStop}%
\bibitem [{\citenamefont {Qi}(2011)}]{Wannier-Qi}%
  \BibitemOpen
  \bibfield  {author} {\bibinfo {author} {\bibfnamefont {X.-L.}\ \bibnamefont
  {Qi}},\ }\href {\doibase 10.1103/PhysRevLett.107.126803} {\bibfield
  {journal} {\bibinfo  {journal} {Phys. Rev. Lett.}\ }\textbf {\bibinfo
  {volume} {107}},\ \bibinfo {pages} {126803} (\bibinfo {year}
  {2011})}\BibitemShut {NoStop}%
\bibitem [{\citenamefont {Taherinejad}\ \emph {et~al.}(2014)\citenamefont
  {Taherinejad}, \citenamefont {Garrity},\ and\ \citenamefont
  {Vanderbilt}}]{WannierSheets}%
  \BibitemOpen
  \bibfield  {author} {\bibinfo {author} {\bibfnamefont {M.}~\bibnamefont
  {Taherinejad}}, \bibinfo {author} {\bibfnamefont {K.~F.}\ \bibnamefont
  {Garrity}}, \ and\ \bibinfo {author} {\bibfnamefont {D.}~\bibnamefont
  {Vanderbilt}},\ }\href {\doibase 10.1103/PhysRevB.89.115102} {\bibfield
  {journal} {\bibinfo  {journal} {Phys. Rev. B}\ }\textbf {\bibinfo {volume}
  {89}},\ \bibinfo {pages} {115102} (\bibinfo {year} {2014})}\BibitemShut
  {NoStop}%
\bibitem [{\citenamefont {Matsumoto}\ \emph {et~al.}(1989)\citenamefont
  {Matsumoto}, \citenamefont {Sasaki},\ and\ \citenamefont
  {Tachiki}}]{Matsumoto1989}%
  \BibitemOpen
  \bibfield  {author} {\bibinfo {author} {\bibfnamefont {H.}~\bibnamefont
  {Matsumoto}}, \bibinfo {author} {\bibfnamefont {M.}~\bibnamefont {Sasaki}}, \
  and\ \bibinfo {author} {\bibfnamefont {M.}~\bibnamefont {Tachiki}},\ }\href
  {\doibase https://doi.org/10.1016/0038-1098(89)90206-8} {\bibfield  {journal}
  {\bibinfo  {journal} {Solid State Communications}\ }\textbf {\bibinfo
  {volume} {71}},\ \bibinfo {pages} {829 } (\bibinfo {year}
  {1989})}\BibitemShut {NoStop}%
\bibitem [{\citenamefont {Haldane}(1988)}]{Haldane}%
  \BibitemOpen
  \bibfield  {author} {\bibinfo {author} {\bibfnamefont {F.~D.~M.}\
  \bibnamefont {Haldane}},\ }\href {\doibase 10.1103/PhysRevLett.61.2015}
  {\bibfield  {journal} {\bibinfo  {journal} {Phys. Rev. Lett.}\ }\textbf
  {\bibinfo {volume} {61}},\ \bibinfo {pages} {2015} (\bibinfo {year}
  {1988})}\BibitemShut {NoStop}%
\bibitem [{\citenamefont {Kane}\ and\ \citenamefont
  {Mele}(2005{\natexlab{a}})}]{Z2_QSH}%
  \BibitemOpen
  \bibfield  {author} {\bibinfo {author} {\bibfnamefont {C.~L.}\ \bibnamefont
  {Kane}}\ and\ \bibinfo {author} {\bibfnamefont {E.~J.}\ \bibnamefont
  {Mele}},\ }\href {\doibase 10.1103/PhysRevLett.95.146802} {\bibfield
  {journal} {\bibinfo  {journal} {Phys. Rev. Lett.}\ }\textbf {\bibinfo
  {volume} {95}},\ \bibinfo {pages} {146802} (\bibinfo {year}
  {2005}{\natexlab{a}})}\BibitemShut {NoStop}%
\bibitem [{\citenamefont {Kane}\ and\ \citenamefont
  {Mele}(2005{\natexlab{b}})}]{Z2_Graphene}%
  \BibitemOpen
  \bibfield  {author} {\bibinfo {author} {\bibfnamefont {C.~L.}\ \bibnamefont
  {Kane}}\ and\ \bibinfo {author} {\bibfnamefont {E.~J.}\ \bibnamefont
  {Mele}},\ }\href {\doibase 10.1103/PhysRevLett.95.226801} {\bibfield
  {journal} {\bibinfo  {journal} {Phys. Rev. Lett.}\ }\textbf {\bibinfo
  {volume} {95}},\ \bibinfo {pages} {226801} (\bibinfo {year}
  {2005}{\natexlab{b}})}\BibitemShut {NoStop}%
\bibitem [{\citenamefont {Kitaev}(2009)}]{Kitaev}%
  \BibitemOpen
  \bibfield  {author} {\bibinfo {author} {\bibfnamefont {A.}~\bibnamefont
  {Kitaev}},\ }\href {\doibase http://dx.doi.org/10.1063/1.3149495} {\bibfield
  {journal} {\bibinfo  {journal} {AIP Conference Proceedings}\ }\textbf
  {\bibinfo {volume} {1134}},\ \bibinfo {pages} {22} (\bibinfo {year}
  {2009})}\BibitemShut {NoStop}%
\bibitem [{\citenamefont {Freed}\ and\ \citenamefont
  {Moore}(2013)}]{FreedMoore}%
  \BibitemOpen
  \bibfield  {author} {\bibinfo {author} {\bibfnamefont {D.~S.}\ \bibnamefont
  {Freed}}\ and\ \bibinfo {author} {\bibfnamefont {G.~W.}\ \bibnamefont
  {Moore}},\ }\href {\doibase 10.1007/s00023-013-0236-x} {\bibfield  {journal}
  {\bibinfo  {journal} {Annales Henri Poincar{\'e}}\ }\textbf {\bibinfo
  {volume} {14}},\ \bibinfo {pages} {1927} (\bibinfo {year}
  {2013})}\BibitemShut {NoStop}%
\bibitem [{\citenamefont {Teo}\ and\ \citenamefont {Kane}(2010)}]{TeoKane}%
  \BibitemOpen
  \bibfield  {author} {\bibinfo {author} {\bibfnamefont {J.~C.~Y.}\
  \bibnamefont {Teo}}\ and\ \bibinfo {author} {\bibfnamefont {C.~L.}\
  \bibnamefont {Kane}},\ }\href {\doibase 10.1103/PhysRevB.82.115120}
  {\bibfield  {journal} {\bibinfo  {journal} {Phys. Rev. B}\ }\textbf {\bibinfo
  {volume} {82}},\ \bibinfo {pages} {115120} (\bibinfo {year}
  {2010})}\BibitemShut {NoStop}%
\bibitem [{\citenamefont {Po}\ \emph {et~al.}(2018{\natexlab{b}})\citenamefont
  {Po}, \citenamefont {Watanabe},\ and\ \citenamefont {Vishwanath}}]{Fragile}%
  \BibitemOpen
  \bibfield  {author} {\bibinfo {author} {\bibfnamefont {H.~C.}\ \bibnamefont
  {Po}}, \bibinfo {author} {\bibfnamefont {H.}~\bibnamefont {Watanabe}}, \ and\
  \bibinfo {author} {\bibfnamefont {A.}~\bibnamefont {Vishwanath}},\ }\href
  {\doibase 10.1103/PhysRevLett.121.126402} {\bibfield  {journal} {\bibinfo
  {journal} {Phys. Rev. Lett.}\ }\textbf {\bibinfo {volume} {121}},\ \bibinfo
  {pages} {126402} (\bibinfo {year} {2018}{\natexlab{b}})}\BibitemShut
  {NoStop}%
\bibitem [{\citenamefont {Marzari}\ \emph {et~al.}(2012)\citenamefont
  {Marzari}, \citenamefont {Mostofi}, \citenamefont {Yates}, \citenamefont
  {Souza},\ and\ \citenamefont {Vanderbilt}}]{VanderbiltRMP}%
  \BibitemOpen
  \bibfield  {author} {\bibinfo {author} {\bibfnamefont {N.}~\bibnamefont
  {Marzari}}, \bibinfo {author} {\bibfnamefont {A.~A.}\ \bibnamefont
  {Mostofi}}, \bibinfo {author} {\bibfnamefont {J.~R.}\ \bibnamefont {Yates}},
  \bibinfo {author} {\bibfnamefont {I.}~\bibnamefont {Souza}}, \ and\ \bibinfo
  {author} {\bibfnamefont {D.}~\bibnamefont {Vanderbilt}},\ }\href {\doibase
  10.1103/RevModPhys.84.1419} {\bibfield  {journal} {\bibinfo  {journal} {Rev.
  Mod. Phys.}\ }\textbf {\bibinfo {volume} {84}},\ \bibinfo {pages} {1419}
  (\bibinfo {year} {2012})}\BibitemShut {NoStop}%
\bibitem [{\citenamefont {Yuan}\ and\ \citenamefont {Fu}(2018)}]{NoahLiang}%
  \BibitemOpen
  \bibfield  {author} {\bibinfo {author} {\bibfnamefont {N.~F.~Q.}\
  \bibnamefont {Yuan}}\ and\ \bibinfo {author} {\bibfnamefont {L.}~\bibnamefont
  {Fu}},\ }\href {\doibase 10.1103/PhysRevB.98.045103} {\bibfield  {journal}
  {\bibinfo  {journal} {Phys. Rev. B}\ }\textbf {\bibinfo {volume} {98}},\
  \bibinfo {pages} {045103} (\bibinfo {year} {2018})}\BibitemShut {NoStop}%
\bibitem [{\citenamefont {Kang}\ and\ \citenamefont {Vafek}(2018)}]{Oskar}%
  \BibitemOpen
  \bibfield  {author} {\bibinfo {author} {\bibfnamefont {J.}~\bibnamefont
  {Kang}}\ and\ \bibinfo {author} {\bibfnamefont {O.}~\bibnamefont {Vafek}},\
  }\href {\doibase 10.1103/PhysRevX.8.031088} {\bibfield  {journal} {\bibinfo
  {journal} {Phys. Rev. X}\ }\textbf {\bibinfo {volume} {8}},\ \bibinfo {pages}
  {031088} (\bibinfo {year} {2018})}\BibitemShut {NoStop}%
\bibitem [{\citenamefont {Koshino}\ \emph {et~al.}(2018)\citenamefont
  {Koshino}, \citenamefont {Yuan}, \citenamefont {Koretsune}, \citenamefont
  {Ochi}, \citenamefont {Kuroki},\ and\ \citenamefont {Fu}}]{KoshinoLiang}%
  \BibitemOpen
  \bibfield  {author} {\bibinfo {author} {\bibfnamefont {M.}~\bibnamefont
  {Koshino}}, \bibinfo {author} {\bibfnamefont {N.~F.~Q.}\ \bibnamefont
  {Yuan}}, \bibinfo {author} {\bibfnamefont {T.}~\bibnamefont {Koretsune}},
  \bibinfo {author} {\bibfnamefont {M.}~\bibnamefont {Ochi}}, \bibinfo {author}
  {\bibfnamefont {K.}~\bibnamefont {Kuroki}}, \ and\ \bibinfo {author}
  {\bibfnamefont {L.}~\bibnamefont {Fu}},\ }\href {\doibase
  10.1103/PhysRevX.8.031087} {\bibfield  {journal} {\bibinfo  {journal} {Phys.
  Rev. X}\ }\textbf {\bibinfo {volume} {8}},\ \bibinfo {pages} {031087}
  (\bibinfo {year} {2018})}\BibitemShut {NoStop}%
\bibitem [{\citenamefont {Xu}\ and\ \citenamefont {Balents}(2018)}]{XuBalents}%
  \BibitemOpen
  \bibfield  {author} {\bibinfo {author} {\bibfnamefont {C.}~\bibnamefont
  {Xu}}\ and\ \bibinfo {author} {\bibfnamefont {L.}~\bibnamefont {Balents}},\
  }\href {\doibase 10.1103/PhysRevLett.121.087001} {\bibfield  {journal}
  {\bibinfo  {journal} {Phys. Rev. Lett.}\ }\textbf {\bibinfo {volume} {121}},\
  \bibinfo {pages} {087001} (\bibinfo {year} {2018})}\BibitemShut {NoStop}%
\bibitem [{\citenamefont {Dodaro}\ \emph {et~al.}(2018)\citenamefont {Dodaro},
  \citenamefont {Kivelson}, \citenamefont {Schattner}, \citenamefont {Sun},\
  and\ \citenamefont {Wang}}]{Dodaro2018}%
  \BibitemOpen
  \bibfield  {author} {\bibinfo {author} {\bibfnamefont {J.~F.}\ \bibnamefont
  {Dodaro}}, \bibinfo {author} {\bibfnamefont {S.~A.}\ \bibnamefont
  {Kivelson}}, \bibinfo {author} {\bibfnamefont {Y.}~\bibnamefont {Schattner}},
  \bibinfo {author} {\bibfnamefont {X.~Q.}\ \bibnamefont {Sun}}, \ and\
  \bibinfo {author} {\bibfnamefont {C.}~\bibnamefont {Wang}},\ }\href {\doibase
  10.1103/PhysRevB.98.075154} {\bibfield  {journal} {\bibinfo  {journal} {Phys.
  Rev. B}\ }\textbf {\bibinfo {volume} {98}},\ \bibinfo {pages} {075154}
  (\bibinfo {year} {2018})}\BibitemShut {NoStop}%
\bibitem [{\citenamefont {Zhang}(2019)}]{Zhang2018}%
  \BibitemOpen
  \bibfield  {author} {\bibinfo {author} {\bibfnamefont {L.}~\bibnamefont
  {Zhang}},\ }\href {\doibase https://doi.org/10.1016/j.scib.2019.03.010}
  {\bibfield  {journal} {\bibinfo  {journal} {Science Bulletin}\ }\textbf
  {\bibinfo {volume} {64}},\ \bibinfo {pages} {495 } (\bibinfo {year}
  {2019})}\BibitemShut {NoStop}%
\bibitem [{\citenamefont {Rademaker}\ and\ \citenamefont
  {Mellado}(2018)}]{Rademaker2018}%
  \BibitemOpen
  \bibfield  {author} {\bibinfo {author} {\bibfnamefont {L.}~\bibnamefont
  {Rademaker}}\ and\ \bibinfo {author} {\bibfnamefont {P.}~\bibnamefont
  {Mellado}},\ }\href {\doibase 10.1103/PhysRevB.98.235158} {\bibfield
  {journal} {\bibinfo  {journal} {Phys. Rev. B}\ }\textbf {\bibinfo {volume}
  {98}},\ \bibinfo {pages} {235158} (\bibinfo {year} {2018})}\BibitemShut
  {NoStop}%
\bibitem [{\citenamefont {Thomson}\ \emph {et~al.}(2018)\citenamefont
  {Thomson}, \citenamefont {Chatterjee}, \citenamefont {Sachdev},\ and\
  \citenamefont {Scheurer}}]{Thomson2018}%
  \BibitemOpen
  \bibfield  {author} {\bibinfo {author} {\bibfnamefont {A.}~\bibnamefont
  {Thomson}}, \bibinfo {author} {\bibfnamefont {S.}~\bibnamefont {Chatterjee}},
  \bibinfo {author} {\bibfnamefont {S.}~\bibnamefont {Sachdev}}, \ and\
  \bibinfo {author} {\bibfnamefont {M.~S.}\ \bibnamefont {Scheurer}},\ }\href
  {\doibase 10.1103/PhysRevB.98.075109} {\bibfield  {journal} {\bibinfo
  {journal} {Phys. Rev. B}\ }\textbf {\bibinfo {volume} {98}},\ \bibinfo
  {pages} {075109} (\bibinfo {year} {2018})}\BibitemShut {NoStop}%
\bibitem [{Note1()}]{Note1}%
  \BibitemOpen
  \bibinfo {note} {Note that $M_y$ may not be present in the actual experiment,
  due to, for instance, different processing of the top and bottom layers.
  Nonetheless, as long as its (explicit) breaking is perturbative, it will be
  helpful to retain it as a theoretical crutch.}\BibitemShut {Stop}%
\bibitem [{Note2()}]{Note2}%
  \BibitemOpen
  \bibinfo {note} {Recall, our ``mirror'' $M_y$ is really a two-fold rotation
  in 3D.}\BibitemShut {Stop}%
\bibitem [{Note3()}]{Note3}%
  \BibitemOpen
  \bibinfo {note} {Or four complex parameters if we lift some fictitious
  symmetry constraints (Appendix \ref {app:Explicit}); one less if we also
  impose normalization.}\BibitemShut {Stop}%
\bibitem [{Note4()}]{Note4}%
  \BibitemOpen
  \bibinfo {note} {We remark that the present construction differs from the
  more conventional approach in which one tries to incorporate the smallest
  number of short-range hoppings to reproduce the salient features in the band
  structure. Instead, our goal here is to illustrate how the introduction of
  quasi-orbitals can naturally produce the nearly flat bands in
  TBG.}\BibitemShut {Stop}%
\bibitem [{\citenamefont {Lieb}(1989)}]{Lieb}%
  \BibitemOpen
  \bibfield  {author} {\bibinfo {author} {\bibfnamefont {E.~H.}\ \bibnamefont
  {Lieb}},\ }\href {\doibase 10.1103/PhysRevLett.62.1201} {\bibfield  {journal}
  {\bibinfo  {journal} {Phys. Rev. Lett.}\ }\textbf {\bibinfo {volume} {62}},\
  \bibinfo {pages} {1201} (\bibinfo {year} {1989})}\BibitemShut {NoStop}%
\bibitem [{Note5()}]{Note5}%
  \BibitemOpen
  \bibinfo {note} {Slightly more technically, these flat bands arise because
  the rank of $h_{\alpha \beta } (\protect \bm {k})$ is at most $\protect
  \qopname \relax m{min}(N_\alpha , N_\beta )$. When $N_\alpha \neq N_\beta $,
  either the left or right null space of $h_{\alpha \beta } (\protect \bm {k})$
  has to be at least $|N_\alpha - N_\beta |$ dimensional for all $\protect \bm
  {k}$. This translates into zero-energy flat bands for $H_{\protect \rm
  Lieb-like}(\protect \bm {k})$.}\BibitemShut {Stop}%
\bibitem [{Note6()}]{Note6}%
  \BibitemOpen
  \bibinfo {note} {Which could split into multiple Dirac cones when trigonal
  warping is incorporated}\BibitemShut {NoStop}%
\bibitem [{\citenamefont {Po}\ \emph {et~al.}(2016)\citenamefont {Po},
  \citenamefont {Watanabe}, \citenamefont {Zaletel},\ and\ \citenamefont
  {Vishwanath}}]{SciAdv}%
  \BibitemOpen
  \bibfield  {author} {\bibinfo {author} {\bibfnamefont {H.~C.}\ \bibnamefont
  {Po}}, \bibinfo {author} {\bibfnamefont {H.}~\bibnamefont {Watanabe}},
  \bibinfo {author} {\bibfnamefont {M.~P.}\ \bibnamefont {Zaletel}}, \ and\
  \bibinfo {author} {\bibfnamefont {A.}~\bibnamefont {Vishwanath}},\ }\href
  {http://advances.sciencemag.org/content/2/4/e1501782} {\bibfield  {journal}
  {\bibinfo  {journal} {Sci. Adv.}\ }\textbf {\bibinfo {volume} {2}} (\bibinfo
  {year} {2016})}\BibitemShut {NoStop}%
\bibitem [{\citenamefont {Po}\ \emph {et~al.}(2017)\citenamefont {Po},
  \citenamefont {Vishwanath},\ and\ \citenamefont {Watanabe}}]{NC}%
  \BibitemOpen
  \bibfield  {author} {\bibinfo {author} {\bibfnamefont {H.~C.}\ \bibnamefont
  {Po}}, \bibinfo {author} {\bibfnamefont {A.}~\bibnamefont {Vishwanath}}, \
  and\ \bibinfo {author} {\bibfnamefont {H.}~\bibnamefont {Watanabe}},\ }\href
  {\doibase 10.1038/s41467-017-00133-2} {\bibfield  {journal} {\bibinfo
  {journal} {Nature Communications}\ }\textbf {\bibinfo {volume} {8}},\
  \bibinfo {pages} {50} (\bibinfo {year} {2017})}\BibitemShut {NoStop}%
\bibitem [{\citenamefont {Bradlyn}\ \emph {et~al.}(2017)\citenamefont
  {Bradlyn}, \citenamefont {Elcoro}, \citenamefont {Cano}, \citenamefont
  {Vergniory}, \citenamefont {Wang}, \citenamefont {Felser}, \citenamefont
  {Aroyo},\ and\ \citenamefont {Bernevig}}]{TopoChem}%
  \BibitemOpen
  \bibfield  {author} {\bibinfo {author} {\bibfnamefont {B.}~\bibnamefont
  {Bradlyn}}, \bibinfo {author} {\bibfnamefont {L.}~\bibnamefont {Elcoro}},
  \bibinfo {author} {\bibfnamefont {J.}~\bibnamefont {Cano}}, \bibinfo {author}
  {\bibfnamefont {M.~G.}\ \bibnamefont {Vergniory}}, \bibinfo {author}
  {\bibfnamefont {Z.}~\bibnamefont {Wang}}, \bibinfo {author} {\bibfnamefont
  {C.}~\bibnamefont {Felser}}, \bibinfo {author} {\bibfnamefont {M.~I.}\
  \bibnamefont {Aroyo}}, \ and\ \bibinfo {author} {\bibfnamefont {B.~A.}\
  \bibnamefont {Bernevig}},\ }\href {http://dx.doi.org/10.1038/nature23268}
  {\bibfield  {journal} {\bibinfo  {journal} {Nature}\ }\textbf {\bibinfo
  {volume} {547}},\ \bibinfo {pages} {298 EP } (\bibinfo {year} {2017})},\
  \bibinfo {note} {article}\BibitemShut {NoStop}%
\bibitem [{\citenamefont {Cano}\ \emph {et~al.}(2018)\citenamefont {Cano},
  \citenamefont {Bradlyn}, \citenamefont {Wang}, \citenamefont {Elcoro},
  \citenamefont {Vergniory}, \citenamefont {Felser}, \citenamefont {Aroyo},\
  and\ \citenamefont {Bernevig}}]{Cano2018}%
  \BibitemOpen
  \bibfield  {author} {\bibinfo {author} {\bibfnamefont {J.}~\bibnamefont
  {Cano}}, \bibinfo {author} {\bibfnamefont {B.}~\bibnamefont {Bradlyn}},
  \bibinfo {author} {\bibfnamefont {Z.}~\bibnamefont {Wang}}, \bibinfo {author}
  {\bibfnamefont {L.}~\bibnamefont {Elcoro}}, \bibinfo {author} {\bibfnamefont
  {M.~G.}\ \bibnamefont {Vergniory}}, \bibinfo {author} {\bibfnamefont
  {C.}~\bibnamefont {Felser}}, \bibinfo {author} {\bibfnamefont {M.~I.}\
  \bibnamefont {Aroyo}}, \ and\ \bibinfo {author} {\bibfnamefont {B.~A.}\
  \bibnamefont {Bernevig}},\ }\href {\doibase 10.1103/PhysRevLett.120.266401}
  {\bibfield  {journal} {\bibinfo  {journal} {Phys. Rev. Lett.}\ }\textbf
  {\bibinfo {volume} {120}},\ \bibinfo {pages} {266401} (\bibinfo {year}
  {2018})}\BibitemShut {NoStop}%
\bibitem [{\citenamefont {{Bouhon}}\ \emph {et~al.}(2018)\citenamefont
  {{Bouhon}}, \citenamefont {{Black-Schaffer}},\ and\ \citenamefont
  {{Slager}}}]{Slager2018}%
  \BibitemOpen
  \bibfield  {author} {\bibinfo {author} {\bibfnamefont {A.}~\bibnamefont
  {{Bouhon}}}, \bibinfo {author} {\bibfnamefont {A.~M.}\ \bibnamefont
  {{Black-Schaffer}}}, \ and\ \bibinfo {author} {\bibfnamefont {R.-J.}\
  \bibnamefont {{Slager}}},\ }\href@noop {} {\  (\bibinfo {year} {2018})},\
  \Eprint {http://arxiv.org/abs/1804.09719} {arXiv:1804.09719} \BibitemShut
  {NoStop}%
\bibitem [{\citenamefont {Fang}\ \emph {et~al.}(2015)\citenamefont {Fang},
  \citenamefont {Chen}, \citenamefont {Kee},\ and\ \citenamefont
  {Fu}}]{Fang2015}%
  \BibitemOpen
  \bibfield  {author} {\bibinfo {author} {\bibfnamefont {C.}~\bibnamefont
  {Fang}}, \bibinfo {author} {\bibfnamefont {Y.}~\bibnamefont {Chen}}, \bibinfo
  {author} {\bibfnamefont {H.-Y.}\ \bibnamefont {Kee}}, \ and\ \bibinfo
  {author} {\bibfnamefont {L.}~\bibnamefont {Fu}},\ }\href {\doibase
  10.1103/PhysRevB.92.081201} {\bibfield  {journal} {\bibinfo  {journal} {Phys.
  Rev. B}\ }\textbf {\bibinfo {volume} {92}},\ \bibinfo {pages} {081201}
  (\bibinfo {year} {2015})}\BibitemShut {NoStop}%
\bibitem [{\citenamefont {Alexandradinata}\ \emph {et~al.}(2016)\citenamefont
  {Alexandradinata}, \citenamefont {Wang},\ and\ \citenamefont
  {Bernevig}}]{Alexandradinata2016}%
  \BibitemOpen
  \bibfield  {author} {\bibinfo {author} {\bibfnamefont {A.}~\bibnamefont
  {Alexandradinata}}, \bibinfo {author} {\bibfnamefont {Z.}~\bibnamefont
  {Wang}}, \ and\ \bibinfo {author} {\bibfnamefont {B.~A.}\ \bibnamefont
  {Bernevig}},\ }\href {\doibase 10.1103/PhysRevX.6.021008} {\bibfield
  {journal} {\bibinfo  {journal} {Phys. Rev. X}\ }\textbf {\bibinfo {volume}
  {6}},\ \bibinfo {pages} {021008} (\bibinfo {year} {2016})}\BibitemShut
  {NoStop}%
\bibitem [{\citenamefont {Ahn}\ \emph {et~al.}(2018)\citenamefont {Ahn},
  \citenamefont {Kim}, \citenamefont {Kim},\ and\ \citenamefont
  {Yang}}]{Ahn2018}%
  \BibitemOpen
  \bibfield  {author} {\bibinfo {author} {\bibfnamefont {J.}~\bibnamefont
  {Ahn}}, \bibinfo {author} {\bibfnamefont {D.}~\bibnamefont {Kim}}, \bibinfo
  {author} {\bibfnamefont {Y.}~\bibnamefont {Kim}}, \ and\ \bibinfo {author}
  {\bibfnamefont {B.-J.}\ \bibnamefont {Yang}},\ }\href {\doibase
  10.1103/PhysRevLett.121.106403} {\bibfield  {journal} {\bibinfo  {journal}
  {Phys. Rev. Lett.}\ }\textbf {\bibinfo {volume} {121}},\ \bibinfo {pages}
  {106403} (\bibinfo {year} {2018})}\BibitemShut {NoStop}%
\bibitem [{\citenamefont {Michel}\ and\ \citenamefont {Zak}(2001)}]{Zak2001}%
  \BibitemOpen
  \bibfield  {author} {\bibinfo {author} {\bibfnamefont {L.}~\bibnamefont
  {Michel}}\ and\ \bibinfo {author} {\bibfnamefont {J.}~\bibnamefont {Zak}},\
  }\href {\doibase http://dx.doi.org/10.1016/S0370-1573(00)00093-4} {\bibfield
  {journal} {\bibinfo  {journal} {Phys. Rep.}\ }\textbf {\bibinfo {volume}
  {341}},\ \bibinfo {pages} {377 } (\bibinfo {year} {2001})}\BibitemShut
  {NoStop}%
\bibitem [{\citenamefont {{Song}}\ \emph {et~al.}(2018)\citenamefont {{Song}},
  \citenamefont {{Wang}}, \citenamefont {{Shi}}, \citenamefont {{Li}},
  \citenamefont {{Fang}},\ and\ \citenamefont {{Bernevig}}}]{Song2018}%
  \BibitemOpen
  \bibfield  {author} {\bibinfo {author} {\bibfnamefont {Z.}~\bibnamefont
  {{Song}}}, \bibinfo {author} {\bibfnamefont {Z.}~\bibnamefont {{Wang}}},
  \bibinfo {author} {\bibfnamefont {W.}~\bibnamefont {{Shi}}}, \bibinfo
  {author} {\bibfnamefont {G.}~\bibnamefont {{Li}}}, \bibinfo {author}
  {\bibfnamefont {C.}~\bibnamefont {{Fang}}}, \ and\ \bibinfo {author}
  {\bibfnamefont {B.~A.}\ \bibnamefont {{Bernevig}}},\ }\href@noop {}
  {\bibfield  {journal} {\bibinfo  {journal} {ArXiv e-prints}\ } (\bibinfo
  {year} {2018})},\ \Eprint {http://arxiv.org/abs/1807.10676}
  {arXiv:1807.10676} \BibitemShut {NoStop}%
\bibitem [{\citenamefont {Nakahara}(2003)}]{Nakahara}%
  \BibitemOpen
  \bibfield  {author} {\bibinfo {author} {\bibfnamefont {M.}~\bibnamefont
  {Nakahara}},\ }\href {https://books.google.com.hk/books?id=cH-XQB0Ex5wC}
  {\emph {\bibinfo {title} {Geometry, Topology and Physics, Second Edition}}},\
  Graduate student series in physics\ (\bibinfo  {publisher} {Taylor \&
  Francis},\ \bibinfo {year} {2003})\BibitemShut {NoStop}%
\end{thebibliography}%

\clearpage

\onecolumngrid
\appendix

\section{Symmetry representations  \label{app:SymRep}}

In this appendix, we providing some details regarding the symmetry representations in the problem.
In Table \ref{tab:orbitals} we provide the symmetry properties of the fermion operators. Note that we have included a time-reversal-like operation $\tilde {\mathcal T}$, which is {\it not} a symmetry of the problem, but is considered for simplifying our discussion.

\begin{table}[h]
\caption{Action of symmetries on the real-space orbitals.
Given a symmetry $g$ and the fermion creation operator $\hat c^\dagger$ for an orbital, we tabulate the outcome of $\hat g \hat c^\dagger \hat g^{-1}$. Note that the action of anti-unitary operators is ambiguous up to an arbitrarily choice on $U(1)$ phases. Also, the time-reversal-like symmetry $\tilde{\mathcal T}$ is {\it not} a symmetry of our problem, as the actual time-reversal symmetry of twisted bilayer graphene exchanges the two valleys.
We include $\tilde{\mathcal T}$ here merely to simplify the discussion, and it would be broken explicitly.
We let $\omega \equiv e^{i 2 \pi /3}$.
\label{tab:orbitals}}
\begin{center}
\begin{tabular}{c|cccc}
\hline \hline
$g$ & $\hat s^\dagger$ & $\hat p_z^\dagger$ & $\hat p_+^\dagger$ & $\hat p_-^\dagger$
 \\
\hline
$C_3$ & $\hat s^\dagger$ & $\hat p_z^\dagger$ & $\omega\, \hat p_+^\dagger$ & $\omega^* \, \hat p_-^\dagger$\\
$M_y$ & $\hat s^\dagger$ & $-\hat p_z^\dagger$  &  $\hat p_-^\dagger$ &  $\hat p_+^\dagger$ \\
$C_2 \mathcal T$ & $\hat s^\dagger$ & $\hat p_z^\dagger$  &  $\hat p_-^\dagger$ &  $\hat p_+^\dagger$ \\
\hline
$\tilde{\mathcal T}$ & $\hat s^\dagger$ & $\hat p_z^\dagger$  &  $- \hat p_-^\dagger$ &  $-\hat p_+^\dagger$ \\
\hline \hline
\end{tabular}
\end{center}
\end{table}

In Table \ref{tab:AIreps}, we provide the momentum-space representations arising from the full-filling of orbitals in real-space, i.e., atomic insulators.
The real-space orbitals we will consider include: ($\tau$, $s$), ($\tau$, $p_z$), ($\tau$, $p_\pm$), ($\eta$, $s$), ($\eta$, $p_z$), ($\eta$, $p_\pm$), ($\kappa$, $s$) and ($\kappa$, $p_z$). Notice that when a pair of states at {$\Gamma$} have $C_3$ representation of ($\omega$, $\omega^*$), these two states must have $M_y$ eigenvalues ($+1$, $-1$). In other words, these two states form the two dimensional representation of symmetry $D_3\backsimeq \langle C_3 \rangle \rtimes \langle M_y \rangle$, where $\langle g \rangle$ indicates the subgroup generated by the element $g$.

\begin{center}
	\begin{table}[h]
		\caption{The resulting symmetry representations at high-symmetry points from various real-space orbitals. Eigenvalues from degenerate bands are grouped by parenthesis. Note that $M_y$ is not a symmetry at ${\rm K}$.
		\label{tab:AIreps}}
			\label{table: tau_s}
		\begin{tabular}{c|cc}
			\hline \hline
			($\tau$, $s$) &
			$\Gamma$ & K\\
			\hline
			$C_3$ & $1$ & $1$\\
			$M_y$ & $1$ & \\
			\hline \hline
		\end{tabular}
		\quad \quad
		\begin{tabular}{c|cc}
			\hline \hline
			($\tau$, $p_z$) &
			$\Gamma$ & K\\
			\hline
			$C_3$ & $1$ & $1$\\
			$M_y$ & $-1$ &\\
			\hline \hline
	 	\end{tabular}
 		\quad \quad
		\begin{tabular}{c|cc}
			\hline \hline
			($\tau$, $p_\pm$) &
			$\Gamma$ & K\\
			\hline
			$C_3$ & ($\omega$, $\omega^*$) & ($\omega$, $\omega^*$)\\
			$M_y$ & $(1, -1)$ &\\
			\hline \hline
		\end{tabular}\\~\\~\\
		
		\begin{tabular}{c|cc}
			\hline \hline
			($\eta$, $s$) &
			$\Gamma$ & K\\
			\hline
			$C_3$ & $1$, $1$ & ($\omega$, $\omega^*$)\\
			$M_y$ & $1$, $1$ &\\
		\hline \hline
		\end{tabular}
		\quad \quad
		\begin{tabular}{c|cc}
			\hline \hline
			($\eta$, $p_z$) &
			$\Gamma$ & K\\
			\hline
			$C_3$ & $1$, $1$ & ($\omega$, $\omega^*$)\\
			$M_y$ & $-1$, $-1$ &\\
			\hline \hline
		\end{tabular}
		\quad \quad
		\begin{tabular}{c|cc}
			\hline \hline
			($\eta$, $p_\pm$) &
			$\Gamma$ & K\\
			\hline
			$C_3$ & ($\omega$, $\omega^*$) , ($\omega$, $\omega^*$) & $1$,  $1$, ($\omega$ , $\omega^*$)\\
			$M_y$ & $(1, -1)$, $(1, -1)$ &\\
			\hline \hline
		\end{tabular}\\~\\~\\
		
	    \begin{tabular}{c|cc}
	    	\hline \hline
	    	($\kappa$, $s$) &
	    	$\Gamma$ & K\\
	    	\hline
	    	$C_3$ & $1$, ($\omega$, $\omega^*$) & $1$, ($\omega$, $\omega^*$)\\
	    	$M_y$ & $1$, $(1,-1)$ &\\
	    	\hline \hline
	    \end{tabular}
	    \quad \quad
        \begin{tabular}{c|cc}
    		\hline \hline
    		($\kappa$, $p_z$) &
    		$\Gamma$ & K\\
    		\hline
    		$C_3$ & $1$, ($\omega$, $\omega^*$) & $1$, ($\omega$, $\omega^*$)\\
    		$M_y$ & $-1$, $(1, -1)$ &\\
    		\hline \hline
    	\end{tabular}
	\end{table}
\end{center}

From Table \ref{tab:AIreps}, one can find all possible representation-matching equations that can be used to resolve the Wannier obstruction in a ``fragile'' manner. 
(Alternatively, one could have also resolved it by appending topological bands, as we discussed in Ref.\ \onlinecite{Jun2018}.)
For example, there are two representation-matching equations involving only three bands:
\begin{equation}
\begin{split}
& (\tau, s) \oplus (\eta, p_z) \overset{{\rm rep.}}{=} (\tau, p_z) \oplus ({\rm target});\\
& (\tau, p_z) \oplus (\eta, s) \overset{{\rm rep.}}{=} (\tau, s) \oplus ({\rm target}).
\end{split}
\end{equation}
This is the minimal number of bands that are needed to resolve the obstruction in terms of representations.
However, as we will see in Appendix \ref{app:Wilson} this resolution cannot correctly capture the band topology of the two nearly flat bands in TBG.
Building from these two representation-matching equations, one can add trivial bands on both sides of the equation and get a new equation. We will call the latter a  derived equation from the former.

When the number of bands involved is four, the representation-matching equations that can potentially lead to a fragile resolution of the obstruction are all derived equations of the above ones. With five bands, however, there are new representation-matching equations:
\begin{equation}
	\begin{split}
		&(\eta, p_z) \oplus (\kappa, s) \overset{{\rm rep.}}{=} (\kappa, p_z) \oplus ({\rm target});\\
		&(\eta, s) \oplus (\kappa, p_z) \overset{{\rm rep.}}{=} (\kappa, s) \oplus ({\rm target}).
	\end{split}
\end{equation}
From these equations and using
\begin{equation} \label{eq: rep-match-kappa}
	\begin{split}
        		(\tau, s) \oplus (\tau, p_\pm) \overset{{\rm rep.}}{=} &(\kappa, s);\\
		(\tau, p_z) \oplus (\tau, p_\pm) \overset{{\rm rep.}}{=} &(\kappa, p_z),
	\end{split}
\end{equation}
one obtains
\begin{equation} \label{eq:rep-5Band}
\begin{split}
& (\eta, p_z) \oplus (\tau, s) \oplus (\tau, p_\pm) \overset{{\rm rep.}}{=} (\kappa, p_z) \oplus ({\rm target}) \overset{{\rm rep.}}{=} (\tau, p_z) \oplus (\tau, p_\pm) \oplus ({\rm target}); \\
& (\eta, s) \oplus (\tau, p_z) \oplus (\tau, p_\pm) \overset{{\rm rep.}}{=} (\kappa, s) \oplus ({\rm target}) \overset{{\rm rep.}}{=} (\tau, s) \oplus (\tau, p_\pm) \oplus ({\rm target}).
\end{split}
\end{equation}

When there are six bands, there are also new representation-matching equations that are not derived ones of equations with fewer bands:
\begin{equation}
\begin{split}
&(\tau, p_z) \oplus (\tau, p_\pm) \oplus (\kappa, s) \overset{{\rm rep.}}{=} (\eta, p_\pm) \oplus ({\rm target});\\
&(\tau, s) \oplus (\tau, p_\pm) \oplus (\kappa, p_z) \overset{{\rm rep.}}{=} (\eta, p_\pm) \oplus ({\rm target}),
\end{split}
\end{equation}
where the first one is precisely Eq. (\ref{eq:RepMatch}) used in the main text. From these equations and using Eq. (\ref{eq: rep-match-kappa}), one can obtain other equations:
\begin{equation}
\begin{split}
(\tau, s) \oplus (\tau, p_z) \oplus (\tau, p_\pm) \oplus (\tau, p_\pm) &\overset{{\rm rep.}}{=} (\eta, p_\pm) \oplus ({\rm target});\\
(\kappa, p_z) \oplus (\kappa, s) &\overset{{\rm rep.}}{=} (\eta, p_\pm) \oplus ({\rm target}).
\end{split}
\end{equation}
Up to six bands, it is straightforward to check that the above are all the independent representation-matching equations that can potentially lead to a fragile resolution of the Wannier obstructions. Here, by ``independent'' we mean that these equations cannot be viewed as a derived one from another.

\section{Details of the Bloch Hamiltonians
\label{app:Explicit}}
In this appendix, we provide further details on the Bloch Hamiltonians constructed in this work. In particular, we provide the explicit expressions for the Hamiltonians.
Unlike the presentation in the main text, we find it more natural to first discuss the six-band model, and then move on to the ten-band one. We will end with a discussion on the five-band model.

\subsection{Six-band model
\label{app:6Band}}
Here, we document explicitly the symmetry-allowed terms entering into the six-band model we constructed. In the following, all the coupling parameters $t$ are real numbers.

Recall our six-band Hilbert space arises from $(\tau, p_z)$, $(\tau, p_\pm)$, and $(\kappa, s)$. Let us write the fermion operator for orbitals in the unit cell $\vec r$ as
\begin{equation}\begin{split}\label{eq:cDef}
\hat {\vec c}^\dagger_{\vec r} \equiv
\left(
\begin{array}{cccccc}
\hat \tau_{p_z; \vec r}^\dagger &
\hat \tau_{p_+; \vec r}^\dagger &
\hat \tau_{p_-; \vec r}^\dagger &
\hat \kappa_{s; \vec r}^{(1)\dagger} &
\hat \kappa_{s; \vec r}^{(2)\dagger} &
\hat \kappa_{s; \vec r}^{(3)\dagger}
\end{array}
\right),
\end{split}\end{equation}
which fixes our basis choice of the  Bloch Hamiltonian.

As discussed in the main text, the terms in our six-band Hamiltonian will mostly be conventional nearest-neighbor bonds, with the sole exception of a second nearest-neighbor bond included for $(\kappa,s)$. Let us discuss these terms one-by-one.
First, the nearest-neighbor bond between the $(\tau,{p_z})$ orbitals takes the standard form on the triangular lattice:
\begin{equation}\begin{split}\label{eq:}
H_{p_z} = &
t_{p_z} \left( \phi_{01} + \phi_{11} + \phi_{10} + {\rm h.c.} \right),
\end{split}\end{equation}
where we let $\phi_{lm} \equiv e^{- i \vec k \cdot (l \vec a_1 + m \vec a_2)}$, and we denote negative numbers by $\bar l \equiv - l$.

The nearest-neighbor couplings for the $(\tau,{p_\pm})$ orbitals are slightly more complicated due to the two-orbital structure. First, the intra-orbital piece is identical to $H_{p_z}$, but with $t_{p_z} \mapsto t_{p_\pm}$ and multiplied by the $2\times 2$ identity matrix. Second, there is also an inter-orbital coupling, which in momentum-space is given by
\begin{equation}\begin{split}\label{eq:}
\mathcal C_{p_\pm p_\pm } =
t_{p_\pm p_\pm }^+ \left( \phi_{01} +\phi_{\bar 1 \bar 1} \omega + \phi_{10} \omega^*\right) +
t_{p_\pm p_\pm }^- \left( \phi_{0 \bar 1} +\phi_{ 1  1} \omega + \phi_{\bar 1 0} \omega^*\right),
\end{split}\end{equation}
where $\omega = e^{i 2 \pi /3}$.
This gives the Bloch Hamiltonian
\begin{equation}\begin{split}\label{eq:}
H_{p_\pm} = &
t_{p_\pm} \left( \phi_{01} + \phi_{11} + \phi_{10} + {\rm h.c.} \right)
\left(
\begin{array}{cc}
1 & 0 \\
0 & 1
\end{array}
\right)
+
\left(
\begin{array}{cc}
0 & C_{p_\pm p_\pm }^\dagger\\
C_{p_\pm p_\pm } & 0
\end{array}
\right).
\end{split}\end{equation}

Next, we consider the kagome lattice. Aside from the standard nearest-neighbor bond $t_{\kappa}$, we find it natural to also incorporate the second nearest-neighbor bond $t_{\kappa}'$, for otherwise there would be an artificial (almost) flat band in the spectrum. This gives
\begin{equation}\begin{split}\label{eq:}
H_\kappa = &
t_\kappa
\left(
\begin{array}{ccc}
0 & \phi_{\bar 1 0}&1 \\
1 & 0 & \phi_{0\bar 1}\\
\phi_{11} & 1 & 0
\end{array}
\right) + t_\kappa'
\left(
\begin{array}{ccc}
0 & \phi_{\bar 1 \bar 1} & \phi_{\bar 1 0 }\\
\phi_{0 \bar 1} & 0 & \phi_{1 0}\\
\phi_{01} & \phi_{11}  & 0
\end{array}
\right) + {\rm h.c.}
\end{split}\end{equation}

\begin{figure}[h]
\begin{center}
{\includegraphics[width=1 \textwidth]{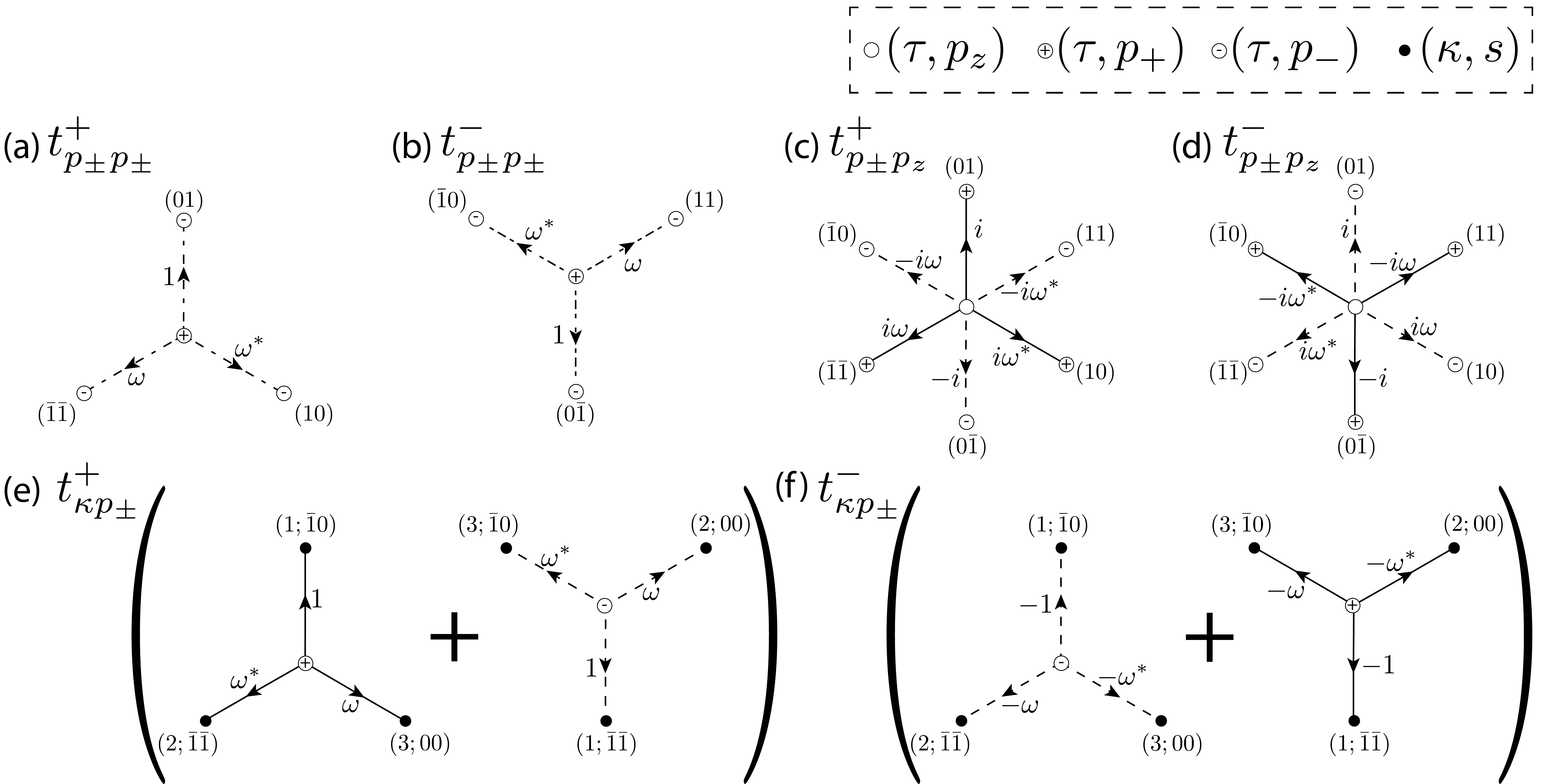}}
\caption{Coupling terms in the six-band model. The full coupling terms consist of the indicated hopping together with their Hermitian conjugates.
We always take the center site to be a triangular site in the ``home'' unit cell, and indicate the unit cell coordinates $l a_1 + m a_2$ of connected sites by $(lm)$ with $\bar l \equiv - l$. For the kagome sites in (e,f), we further specify their sublattice indices.
The strength of the terms in panels (a-f) are respectively denoted by
$t_{p_\pm  p_\pm }^+$, $t_{p_\pm  p_\pm }^-$, $t_{p_\pm p_z}^+$, $t_{p_\pm p_z}^-$,  $t_{\kappa p_\pm}^+$, and  $t_{\kappa p_\pm}^-$.
\label{fig:Bonds6}
 }
\end{center}
\end{figure}

Lastly, we consider the nearest-neighbor coupling between the different lattices.
In momentum space they are characterized by
\begin{equation}\begin{split}\label{eq:}
\mathcal C_{p_\pm p_z}
= &
i t^+_{p_\pm p_z}
\left(
\begin{array}{c}
\phi_{01} +  \phi_{\bar 1 \bar 1} \, \omega +  \phi_{10}\, \omega^* \\
- \left(
\phi_{0 \bar 1} +  \phi_{11}\, \omega^*+  \phi_{\bar 1 0}\, \omega
\right)
\end{array}
\right)
- i t^-_{p_\pm p_z}
\left(
\begin{array}{c}
\phi_{0 \bar 1} +  \phi_{11}\, \omega+  \phi_{\bar 1 0}\, \omega^*
\\
- \left(\phi_{01} +  \phi_{\bar 1 \bar 1}\,  \omega^* +  \phi_{10} \,\omega\right)
\end{array}
\right);\\
\mathcal C_{\kappa p_\pm}
= &
t^+_{\kappa p_\pm}
\left(
\begin{array}{cc}
\phi_{\bar 1 0} & \phi_{\bar 1 \bar 1}\\
\phi_{\bar 1 \bar 1}\, \omega^*  & \omega\\
\omega & \phi_{\bar 1 0} \, \omega^*
\end{array}
\right)
- t^-_{\kappa p_\pm}
\left(
\begin{array}{cc}
 \phi_{\bar 1 \bar 1} & \phi_{\bar 1 0}\\
\omega^* & \phi_{\bar 1 \bar 1}\, \omega  \\
 \phi_{\bar 1 0} \, \omega &\omega^*
\end{array}
\right).
\end{split}\end{equation}
Here, the subscript $p_\pm p_z$ indicates coupling between the $(\tau,{p_{\pm}})$ and the $(\tau, {p_z})$ orbitals, and $\kappa p_\pm$ indicates that between $(\kappa,{s})$ and $(\tau,{p_{\pm}})$.
The real-space form of all the nontrivial couplings above is represented diagrammatically in Fig.\ \ref{fig:Bonds6}.

Altogether, the full Bloch Hamiltonian of the six-band model is given by
\begin{equation}\begin{split}\label{eq:}
H^{(6)}_{\vec k} =
\left(
\begin{array}{ccc}
H_{p_z} +\mu_{p_z} & \mathcal C_{p_\pm p_z}^\dagger & 0\\
\mathcal C_{p_\pm p_z} & H_{p_\pm} +\mu_{p_\pm} & \mathcal C_{\kappa p_\pm}^\dagger \\
0 & \mathcal C_{\kappa p_\pm} & H_{\kappa} +\mu_\kappa
\end{array}
\right)
\end{split}\end{equation}
where we have added relative chemical potentials $\mu_i$ across the different lattices.
For reasons that will become clearer later, we find it convenient to reparameterize them as
\begin{equation}\begin{split}\label{eq:}
\mu_{p_z} \equiv -6 t_{p_z} + \delta_{p_z};~~~~~\mu_{p_\pm} \equiv 3 t_{p_\pm} + \delta_{p_\pm}; ~~~ \&~~~\mu_\kappa \equiv -4(t_\kappa + t_\kappa') + \delta_\kappa.
\end{split}\end{equation}

Before we proceed, we make two remarks regarding the model parameters.
First, note that we have ignored the coupling between the $(\tau,{p_z})$ and $(\kappa,{s})$ orbitals, as we find its inclusion to be unnecessary for reproducing the key energetic features of the dispersion. In practice, such terms are symmetry-allowed and would never be exactly zero, but since we only address symmetry-robust features in the problem their presence has little physical implications.
As such, we choose to keep it at $0$ to simplify the discussion.

Second, we have parameterized the coupling strengths such that when $t_{p_\pm p_\pm }^+  - t_{p_\pm p_\pm }^- = t_{p_\pm p_z}^+ - t_{p_\pm p_z}^- = t_{\kappa p_\pm}^+ - t_{\kappa p_\pm}^- = 0$, $H^{(6)}$ will be $\tilde{\mathcal T}$-invariant. Since the time-reversal-like operation $\tilde{\mathcal T}$ is not an actual symmetry of the problem, this relation will not hold in our model parameters. This parameterization is nonetheless adopted as it provides a simple way to control the degree of $\tilde{\mathcal T}$-breaking in the spectrum.

Next, we expand on the discussion in the main text concerning a physical picture for our model parameters. Recall that the electronic weights of TBG sit mostly at the AA spots, which form a triangular lattice at the \moire scale \cite{Magaud2010, Uchida2014, PabloPRL, ShiangPRB,STMPRL, STMNatPhy,Crommie2015}. In addition, the Dirac points at K and K' are naturally explained by the symmetry characters of the $(\tau, {p_{\pm}})$ orbitals.
The main nontrivial feature of the nearly flat bands, therefore, stems from the fact that at $\Gamma$ points the non-degenerate bands {\it cannot} come from the $(\tau, {p_{\pm}})$ orbitals due to a mismatch in the symmetry representations. Rather, in our Hilbert space they can only arise from the $(\tau,{p_z})$ and $(\kappa,{s})$ orbitals. Our goal is to choose parameters such that the orbital characters of the nearly flat bands behave as expected across the entire Brillouin zone (BZ).

First, let us ignore all the nontrivial coupling terms by setting $t_{p_\pm p_\pm }^\pm = t_{p_\pm p_z}^\pm = t_{\kappa p_\pm}^\pm = 0$.
As discussed, the nearly flat bands will be formed by the $(\tau,{p_z})$ and $(\kappa,{s})$ orbitals at $\Gamma$, and the $(\tau,{p_{\pm}})$ Dirac points at K and K'.
In our parameterization, we can arrange all these orbitals to be at zero-energy by setting $\delta_{p_z} = \delta_{p_\pm} = \delta_{\kappa} = 0$. Furthermore, by adjusting the values of $\delta_{p_z}$ and $\delta_{\kappa}$ we can set the bandwidth of the nearly flat bands.

As the wave functions of the nearly flat bands at $\Gamma$ must arise from the $(\tau, p_z)$ and $(\kappa,s)$ orbitals, we should bring down the energy of the $(\tau,{p_{\pm}})$ bands at $\Gamma$. This is achieved by choosing $t_{p_\pm} <0$.
This sets up the required orbital characters for the bands near zero energy: 
$(\tau, p_z)$ and $(\kappa,s)$  at $\Gamma$ but $(\tau, p_\pm)$ in the rest of the BZ.
We are then left with the $(\tau,{p_z})$ and $(\kappa,{s})$ bands at most of the BZ, as well as the $(\tau,{p_{\pm}})$-Dirac point at $\Gamma$.
Observe that the bandwidth of the four connected bands below the nearly flat bands set the dominant energy scale in the band manifold we intend to describe. This bandwidth can be reconciled with the size of $t_{\kappa}$, and therefore we choose it to be the reference energy scale and measure all the other terms with respect to it. We also choose the values of $t_{p_z}$ and $t_{\kappa}'$ to reproduce the broad energetic features of the bands.

Fig.\ \ref{fig:6Band}b in the main text shows the band dispersion we obtained following the discussion above. Observe that the broad energetic features are already in place. It remains to turn on the nontrivial couplings between the orbitals, and open band gaps to isolate the two nearly flat bands at the top. Our chosen parameters are tabulated in Table \ref{tab:Para6Band}, and the corresponding band structure is shown in Fig.\ \ref{fig:6Band}c in the main text. Note that some parameters are roughly an order of magnitude smaller than the others; they are adjusted to capture the fine energetic features in the nearly flat bands.

\begin{center}
\begin{table}[h]
\caption{Hopping parameters in the six-band model. We abbreviate ``nearest neighbor'' to ``nnbr.'' We set the dominant energy scale to be $t_\kappa = 27$ meV, and measure all the other terms relative to that.
\label{tab:Para6Band}}
\begin{tabular}{c|c|c}
\hline \hline
Parameter  & Meaning & Ratio to $t_{\kappa}$\\
\hline
$\delta_{p_z} \equiv \mu_{p_z} + 6 t_{p_z}$ & $(\tau,{p_z})$ chemical potential & $0$\\
$\delta_{p_\pm} \equiv \mu_{p_\pm}- 3 t_{p_\pm} $ & $(\tau,{p_\pm})$ chemical potential & $-0.23$\\
$\delta_{\kappa} \equiv \mu_\kappa + 4(t_\kappa + t_\kappa') $ & $(\kappa, {s})$ chemical potential & $0.25$\\
$t_{p_z}$ & $(\tau,{p_z})$ nnbr & $0.17$\\
$t_{p_\pm}$ & $(\tau,{p_\pm})$ intra-orbital nnbr & $-0.017$\\
$t_{p_\pm p_\pm }^+$ & $(\tau,{p_\pm})$ inter-orbital nnbr & $-0.065$\\
$t_{p_\pm p_\pm }^-$ & $(\tau,{p_\pm})$ inter-orbital nnbr & $-0.055$\\
$t_{\kappa}$ & $(\kappa,{s})$ nnbr & $1$\\
$t_{\kappa}'$ &$(\kappa,{s})$ second-nnbr & $0.25$\\
$t_{p_\pm p_z}^+$ & $(\tau,{p_\pm})$-$(\tau,{p_z})$ nnbr & $0.095$\\
$t_{p_\pm p_z}^-$ & $(\tau,{p_\pm})$-$(\tau,{p_z})$ nnbr & $0.055$\\
$t_{\kappa p_\pm}^+$ & $(\kappa,{s})$-$(\tau,{p_\pm})$ nnbr & $0.6$\\
$t_{\kappa p_\pm}^-$ & $(\kappa,{s})$-$(\tau,{p_\pm})$ nnbr & $0.2$\\
\hline \hline
\end{tabular}
\end{table}
\end{center}

\subsection{Ten-band model
\label{app:10Band}}
Next, we discuss the terms in the ten-band model. Here, the dominant energetic features are imprinted by the choice of the ``quasi-orbital'' fermion operator $\hat h_{p_+;\vec r}^{(l) \dagger} \equiv \sum_{\vec x}\hat c^\dagger_{\vec x} h_{p_+;\vec r}^{(l)}( \vec x)$. We have already provided the explicit form of the localized wave functions $h_{p_+;\vec r}^{(l)}( \vec x)$ in real space in Fig.\ \ref{fig:Sites}b of the main text. However, it will be convenient to also write down explicitly the Fourier transform of these wave function in momentum space:
\begin{equation}\begin{split}\label{eq:}
h_{p_+;\vec k}^{(A)} =
\left(
\begin{array}{c}
- \left( \omega + \phi_{11} \omega^* + \phi_{01}\right) \zeta^* a\\
\left( \omega^* + \phi_{11} \omega + \phi_{01} \right) \zeta b\\
\left(1+  \phi_{11} + \phi_{01}  \right) c\\
- i \phi_{\bar 1 0} d\\
- i \omega \,d\\
- i \phi_{01} \omega^* d
\end{array}
\right);
~&~
h_{p_-;\vec k}^{(A)}=
\left(
\begin{array}{c}
\left( 1 + \phi_{11} \omega^* + \phi_{01}\omega \right)  \zeta^* a\\
\left( 1 + \phi_{11}  + \phi_{01} \right) c\\
\left(1+  \omega \phi_{11} + \omega^* \phi_{01}  \right) \zeta b\\
-i \phi_{\bar 1 0} d\\
-i \omega^* d\\
-i \phi_{01} \omega \,d
\end{array}
\right);\\
h_{p_+;\vec k}^{(B)} =
\left(
\begin{array}{c}
\left( \omega + \phi_{10} \omega^* + \phi_{11}\right) \zeta a\\
\left( \omega^* + \phi_{10} \omega + \phi_{11} \right) \zeta^* b\\
\left(1+  \phi_{10} + \phi_{11}  \right) c\\
i d\\
i \omega \,d\\
i \omega^* d
\end{array}
\right);
~&~
h_{p_-;\vec k}^{(B)}=
\left(
\begin{array}{c}
-\left( \omega + \phi_{10}  + \phi_{11} \omega^* \right) \zeta a\\
\left( 1 + \phi_{10}  + \phi_{11} \right) c\\
\left( \omega^*+   \phi_{10} + \phi_{11} \omega \right) \zeta^* b\\
i d\\
i\omega^* d\\
i\omega\, d
\end{array}
\right).
\end{split}\end{equation}
Furthermore, it is natural to group these four column vectors into a single $6\times 4$ matrix:
\begin{equation}\begin{split}\label{eq:}
h_{\vec k} = \left(
\begin{array}{cccc}
h_{p_+;\vec k}^{(A)} & h_{p_-;\vec k}^{(A)} & h_{p_+;\vec k}^{(B)} & h_{p_-;\vec k}^{(B)}
\end{array}
\right).
\end{split}\end{equation}
The Bloch Hamiltonian corresponding to Eq.\ \eqref{eq:10Band} with $\delta = 0$ can then be written as
\begin{equation}\begin{split}\label{eq:}
H_{\vec k}(t,0) =
t \left(
\begin{array}{cc}
0_{6\times 6} & h_{\vec k} \\
h_{\vec k}^\dagger & 0_{4\times 4}
\end{array}
\right).
\end{split}\end{equation}
Note the block structure of $H(t,0)$; we see immediately that $H(t,0)$ anti-commutes with $\openone_{6\times 6} \oplus (-\openone_{4\times 4} )$, implying $H_{\vec k}(t,0)$ will be particle-hole symmetric at every $\vec k$. Also, as $h_{\vec k}^\dagger$ is a $4\times 6$ rectangular matrix, the equation
\begin{equation}\begin{split}\label{eq:}
h_{\vec k}^\dagger \varphi = 0
\end{split}\end{equation}
must have at least two non-trivial solutions at every $\vec k$. In other words, there will be (at least) two exactly flat bands at zero-energy in the spectrum of $H(t,0)$ (Fig.\ \ref{fig:Bands}a and d in the main text). For generic choices of the parameters $a$--$d$, these will be the only states at zero-energy, and they serve as the precursor for the nearly flat bands in the final model.

\begin{center}
\begin{table}[h]
\caption{Hopping parameters in the perturbation $\hat V$ to the ten-band model. We abbreviate ``nearest neighbor'' to ``nnbr.'' 
We measure the strengths of the various terms relative to the dominate one, $t_\eta = 32.5$ meV.
All terms present in Table \ref{tab:Para6Band} but not here are set to $0$.
\label{tab:Para10Band}}
\begin{tabular}{c|c|c}
\hline \hline
Parameter  & Meaning & Ratio to $t_\eta$\\
\hline
$\delta_{p_z} \equiv \mu_{p_z} + 6 t_{p_z}$ & $(\tau,{p_z})$ chemical potential & $-0.100$\\
$\delta_{\kappa} \equiv \mu_\kappa + 4(t_\kappa + t_\kappa') $ & $(\kappa,{s})$ chemical potential & $0.110$\\
$t_{p_z}$ & $(\tau,{p_z})$ nnbr & $0$\\
$t_{p_\pm}$ & $(\tau,{p_\pm})$ intra-orbital nnbr & $0.003$\\
$t_{p_\pm p_\pm }^-$ & $(\tau,{p_\pm})$ inter-orbital nnbr & $0.004$\\
$t_{\kappa}$ & $(\kappa,{s})$ nnbr & $0$\\
$t_{\kappa}'$ &$(\kappa,{s})$ second-nnbr & $0$\\
$t_{p_\pm p_z}^+$ & $(\tau,{p_\pm})$-$(\tau,{p_z})$ nnbr & $0.016$\\
$t_{\kappa p_\pm}^+$ & $(\kappa,{s})$-$(\tau,{p_\pm})$ nnbr & $0.016$\\
$t_{\kappa p_\pm}^-$ & $(\kappa,{s})$-$(\tau,{p_\pm})$ nnbr & $-0.016$\\
\hline
$ t_\eta \, e^{i\phi_\eta}$ & $(\eta,{p_\pm})$ nnbr & $i$\\
\hline \hline
\end{tabular}
\end{table}
\end{center}

\begin{figure}[h!]
\begin{center}
{\includegraphics[width=0.55 \textwidth]{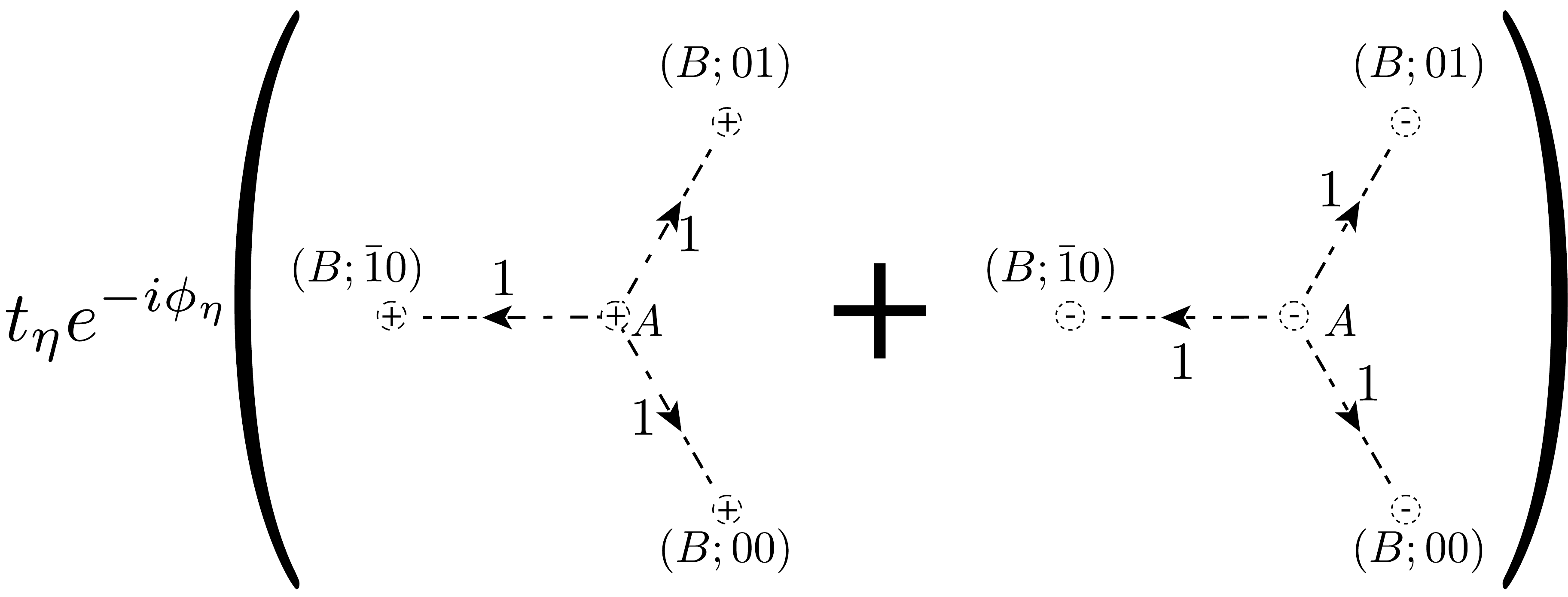}}
\caption{The leading perturbation to the ten-band model, which breaks the undesirable $\tilde{\mathcal T}$-invariance. Each circle denotes a {\it honeycomb} site, with the orbital $p_{\pm}$ indicated.
\label{fig:Heta}
 }
\end{center}
\end{figure}

Our next step is to add generic perturbations to reproduce the actual energetic features of the TBG spectrum. There are two main features which we wish to capture: (i) the dispersion of the nearly flat bands; and (ii) the absence of $\tilde {\mathcal T}$ in the higher-energy bands. For (i), we simply add a subset of the terms we used in constructing the six-band model.
For (ii), we consider an additional nearest-neighbor coupling between the $(\eta, p_{\pm})$ orbitals, which are present only in the ten-band Hilbert space. As in the previous discussion, we represent the term diagrammatically in Fig.\ \ref{fig:Heta}, and provide the explicit expression:
\begin{equation}\begin{split}\label{eq:}
H_\eta =
t_\eta
\left(
\begin{array}{cc}
0 & e^{i \phi_\eta} (1 +   \phi_{0 \bar 1} + \phi_{ 1 0}) \\
e^{-i \phi_\eta} (1 + \phi_{01} + \phi_{\bar 1 0})& 0
\end{array}
\right) \otimes \openone_{2\times 2},
\end{split}\end{equation}
where  $t_\eta, \phi_\eta$ are real parameters. In order to break the undesirable $\tilde{\mathcal T}$-invariance, we will choose $\phi_{\eta} \neq 0,\pi$. 

Altogether, the perturbation to the ten-band model is, in a block-matrix form, given by
\begin{equation}\begin{split}\label{eq:}
V_{\vec k} =
\left(
\begin{array}{cccc}
\mu_{p_z} & \mathcal C_{p_\pm p_z}^\dagger & 0 &0\\
\mathcal C_{p_\pm p_z} & H_{p_\pm} +\mu_{p_\pm} & \mathcal C_{\kappa p_\pm}^\dagger &0\\
0 & \mathcal C_{\kappa p_\pm} & \mu_\kappa & 0\\
0 & 0 & 0 & H_\eta
\end{array}
\right),
\end{split}\end{equation}
where the four blocks correspond respectively to the $(\tau, p_z)$,  $(\tau, p_\pm)$, $(\kappa, s)$, and $(\eta, p_{\pm})$ orbitals. The chosen parameters for the relative strengths of the terms are provided in Table \ref{tab:Para10Band}.
Note that, if we switch off all the perturbations other than $H_\eta$, the two bands at charge neutrality will remain exactly flat.

\subsection{Five-band model
\label{app:5Band}}
Here, we provide the Bloch Hamiltonian for the five-band model defined in Sec.\ \ref{sec:5Band} of the main text. Similar to our discussion of the ten-band Hamiltonian, the model is formulated in terms of the ``quasi-orbitals'' $\rho^{(l)}_{s;\vec r}$ defined in Fig.\ \ref{fig:Sites5}. Under Fourier transform, one finds
\begin{equation}\begin{split}\label{eq:}
\rho_{s;\vec k}^{(1)} = 
\left(
\begin{array}{c}
i \tilde{a} (\phi_{11} - \phi_{10})\\
\tilde{b}\,\phi_{11} +\tilde{c}\,\phi_{10}\\
\tilde{c}\,\phi_{11} +\tilde{b}\,\phi_{10}\\
\tilde{d}^*\,\phi_{10}\\
\tilde{d}
\end{array}
\right);~~~
\rho_{s;\vec k}^{(2)} = 
\left(
\begin{array}{c}
i \tilde{a} (1 - \phi_{11})\\
\omega (\tilde{b} +\tilde{c}\,\phi_{11}) \\
\omega^*(\tilde{c}+\tilde{b}\,\phi_{11})\\
\tilde{d}^*\\
\tilde{d}
\end{array}
\right);~~~
\rho_{s;\vec k}^{(3)} = 
\left(
\begin{array}{c}
i \tilde{a} (\phi_{10} - 1)\\
\omega^*(\tilde{b}\,\phi_{10} +\tilde{c} )\\
\omega(\tilde{c}\,\phi_{10} +\tilde{b})\\
\tilde{d}^*\,\phi_{0 \bar 1}\\
\tilde{d}
\end{array}
\right),
\end{split}\end{equation}
where $\tilde{a}, \tilde{b}, \tilde{c}$ are real and $\tilde{d}$ can be complex.
Going from top to bottom, the entries correspond to the fermion operators $\hat \tau_{p_z; \vec k}$, $\hat \tau_{p_+; \vec k}$,$\hat \tau_{p_-; \vec k}$, $\hat \eta_{s; \vec k}^{(A)}$, and $\hat \eta_{s; \vec k}^{(B)}$.
Furthermore, we again aggregate the three column vectors into a $5\times 3$ matrix $\rho_{\vec k}\equiv \left(
\begin{array}{ccc}
\rho_{s;\vec k}^{(1)} & \rho_{s;\vec k}^{(2)}& \rho_{s;\vec k}^{(3)}
\end{array}
\right)$. Then we can write the Bloch Hamiltonian as
\begin{equation}\begin{split}\label{eq:}
H^{(5)}_{\vec k} = -  t_0' \rho_{\vec k} \rho_{\vec k}^\dagger + {\rm diag}(\mu_{p_z}',\mu_{p_\pm}',\mu_{p_\pm}',\mu_{\eta}',\mu_{\eta}'),
\end{split}\end{equation}
where the $\mu'_i$ are again chemical potential.
Again, if we set $\mu'_i=0$, there will be two exactly flat bands at zero energy, and, conversely, one can reproduce the energetic features of the two nearly flat bands simply by adjusting the chemical potentials $\mu'_i$.

\section{Deformation to explicit atomic limits for the six- and five-band models
\label{app:Deform}}
For completeness, we demonstrate the triviality of the complementary bands in the six- and five-band models in this appendix.

Unlike the ten-band model, our six-band model was defined using only conventional hopping terms without resorting to the notion of quasi-orbitals. Consequently, the four complementary bands are not automatically trivial. This can be settled by a similar deformation to an explicit atomic limit, achieved by first augmenting the Hilbert space to include the $(\eta, p_\pm)$ bands, and then writing down a deformation Hamiltonian akin to that in Eq.\ \eqref{eq:Deform} in the main text:
\begin{equation}\begin{split}\label{eq:Deform6Band}
H_\mu^{'(6)} = t_0 \cos\left( \frac{\pi \mu}{2 \mu_0}\right)\left(
\begin{array}{cc}
0 & \tilde h_{\vec k}\\
\tilde h_{\vec k}^\dagger & 0
\end{array}
\right)
+ \mu
\left(
\begin{array}{cc}
\frac{1}{\mu_0}H^{(6)}_{\vec k} & 0\\
0 & \frac{1}{10} \openone
\end{array}
\right).
\end{split}\end{equation}
Here, $\tilde h_{\vec k}$ is identical to that defined in Appendix \ref{app:10Band}, but with a different set of wave-function parameters
$(a,b,c,d) = (0.48 - 0.24 i, 0.13 + 0.42 i, 0.04 + 0.30 i, 0.24 - 0.29 i)$. As before, we set $t_0 = \mu_0 = 130$ meV.

\begin{figure}[h]
\begin{center}
{\includegraphics[width=0.7 \textwidth]{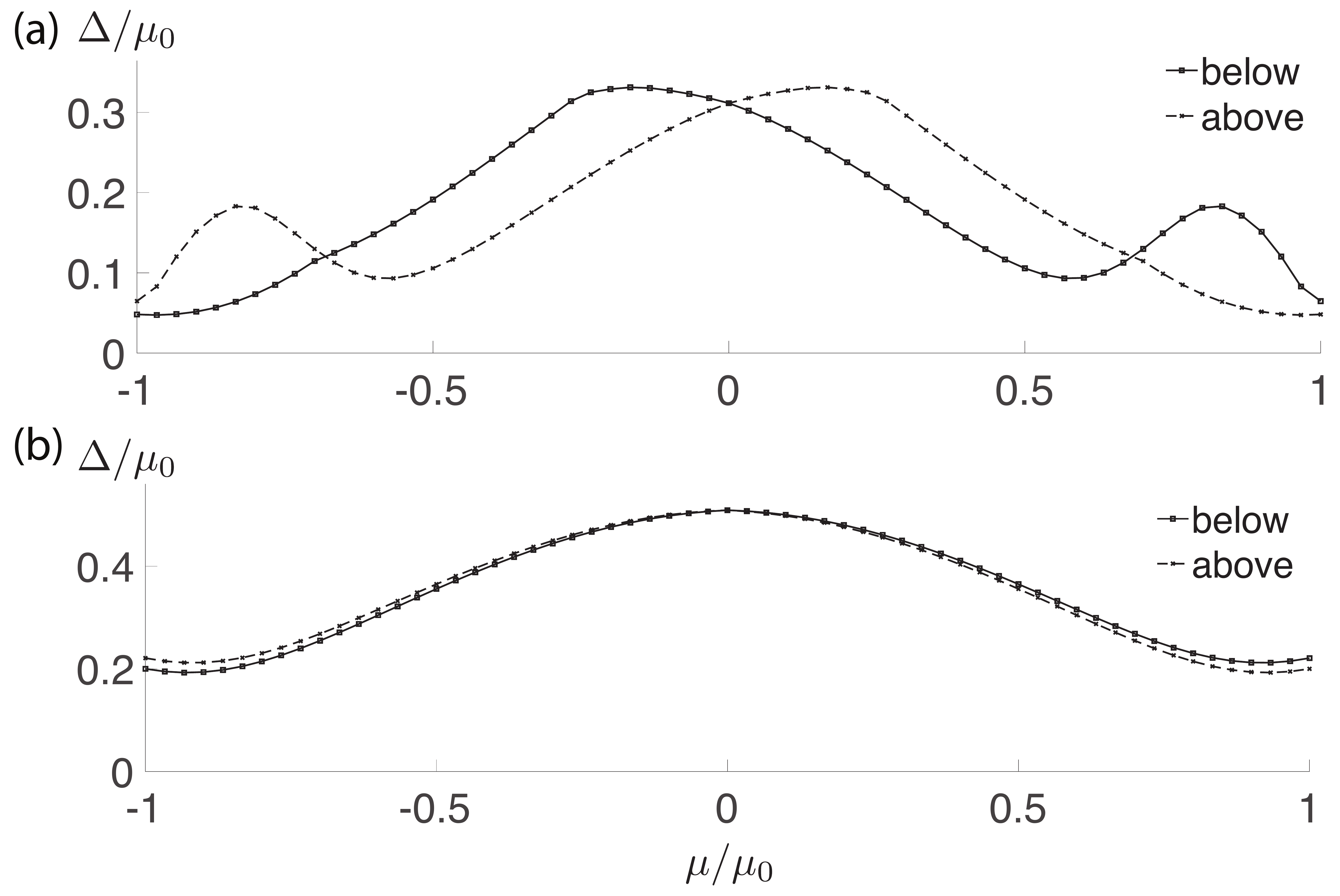}}
\caption{
Evolution of the band gaps above and below the two nearly flat bands in the deforming Hamiltonians for (a) the six-band model in Eq.\ \eqref{eq:Deform6Band}, and (b) the five-band model in Eq.\ \eqref{eq:Deform5Band}. This establishes an adiabatic deformation of the complementary bands in these models to a strict atomic insulator.
More than $4 \times 10^4$ momenta are sampled in the BZ in determining $\Delta$.
\label{fig:Deform6Band}
 }
\end{center}
\end{figure}

When $\mu = \mu_0$, the lowest four bands coincide with that of $H^{(6)}$; when $\mu = - \mu_0$, the lowest four bands correspond to the strict atomic insulator arising from the $(\eta, p_\pm)$ orbitals.
In Fig.\ \ref{fig:Deform6Band}a, we plot the evolution of the band gaps above and below the two nearly flat bands at charge neutrality. The gaps never close, and hence $H_\mu^{'(6)}$ represents a smooth deformation of the lowest four bands of $H^{(6)}$ to a strict atomic insulator.

Next, let us show that the three complementary bands in our five-band model are  atomic in nature. This is again anticipated, as the bands are constructed through the notion of quasi-orbitals transforming in the same way as $(\kappa,s)$. We will demonstrate it explicitly in the same manner as it was done for the six-band model:
 We first augment the Hilbert space to introduce explicitly a set of $(\kappa,s)$ orbitals, and then consider the deformation Hamiltonian

\begin{equation}\begin{split}\label{eq:Deform5Band}
H_\mu^{'(5)} =  t_D' \cos\left( \frac{\pi \mu}{2 \mu_0}\right)\left(
\begin{array}{cc}
0 &  \rho_{\vec k}\\
\rho_{\vec k}^\dagger & 0
\end{array}
\right)
+ \mu
\left(
\begin{array}{cc}
\frac{1}{\mu_0}H^{(5)}_{\vec k} & 0\\
0 & \frac{1}{4} \openone
\end{array}
\right),
\end{split}\end{equation}
here, all the parameters are the same as those listed in the caption of Fig.\ \ref{fig:5Bands}, and we set $\mu_0 = t_D'= t_0' = 80$ meV. The evolution of the band gaps above and below the two nearly flat bands are shown in Fig.\ \ref{fig:Deform6Band}b. As with the other models, the gaps never close, which establishes an adiabatic deformation of the three complementary bands to a strict atomic limit.

\section{``Wilson loop'' analysis
\label{app:Wilson}}
In Ref.\ \onlinecite{Song2018}, it was proposed that the band topology of the two nearly flat bands of TBG could be inferred from a certain topological property of the Wilson loop, and the authors further suggested that this band topology is ``stable,'' in that it will survive even in the presence of additional trivial degrees of freedom.

In essence, ``Wilson loops'' are the multi-band generalization of the Berry phase.
To be self-contained, we define it as follows:
Consider a set of $n$ bands whose eigenvectors are collected into a matrix $\Psi_{\vec k}$ for every $\vec k$.
Let $\mathcal C$ be a closed path in the BZ, and let $\{ \vec k_i\}$ be a discretization of $\mathcal C$ into $N$ momenta such that $|\vec k_{i+1} - \vec k_{i}| \rightarrow 0$ as $N\rightarrow \infty$.
We further label the momenta such that. $\vec k_{N} = \vec k_{1}+ \vec b $, where $\vec b$ (possibly $= \vec 0$) is a reciprocal lattice vector encoding the topological property of $\mathcal C$ (as a loop over the BZ).
Then we define the Wilson loop to be the $n\times n$ matrix
\begin{equation}\begin{split}\label{eq:}
\mathcal W(\mathcal C) \equiv& \lim_{N\rightarrow \infty} \Psi_{\vec k_1}^\dagger
 \Psi_{\vec k_2}  \Psi_{\vec k_2}^\dagger \dots
  \Psi_{\vec k_{N-1}}  \Psi_{\vec k_{N-1}}^\dagger \Psi_{\vec k_N},
\end{split}\end{equation}
where care must be taken to relate $\Psi_{\vec k_N} = \mathcal U_{\vec k_1, \vec b} \Psi_{\vec k_1}$ for some unitary matrix $\mathcal U_{\vec k_1, \vec b}$. $\mathcal W(\mathcal C)$ is unitary when $N\rightarrow \infty$, and in the presence of $\mathcal C_2 \mathcal T$ symmetry it can be further shown to be orthogonal \cite{Fang2015, Alexandradinata2016, Ahn2018, Slager2018, Song2018}.

Following the recipe in Ref.\ \onlinecite{Song2018}, we compute $\mathcal W(\mathcal C)$ for a particular set of contours: write any $\vec k$ in the BZ as $\vec k = \frac{k_\parallel}{2\pi} \vec b_\parallel +\frac{k_\perp }{2\pi} \vec b_\perp$, where $\vec b_\parallel $ and $\vec b_\perp$ are distinct reciprocal lattice vectors. Then pick $\mathcal C$ to be ``straight lines'' running along $\vec b_\parallel$ wrapping around the BZ once. The different contours are then labeled by the remaining coordinate $k_\perp$, and one studies the properties of the family of Wilson loops $\{\mathcal W(k_\perp) ~:~ k_\perp \in [0,2 \pi)\}$.

\begin{figure}[h]
\begin{center}
{\includegraphics[width=0.75 \textwidth]{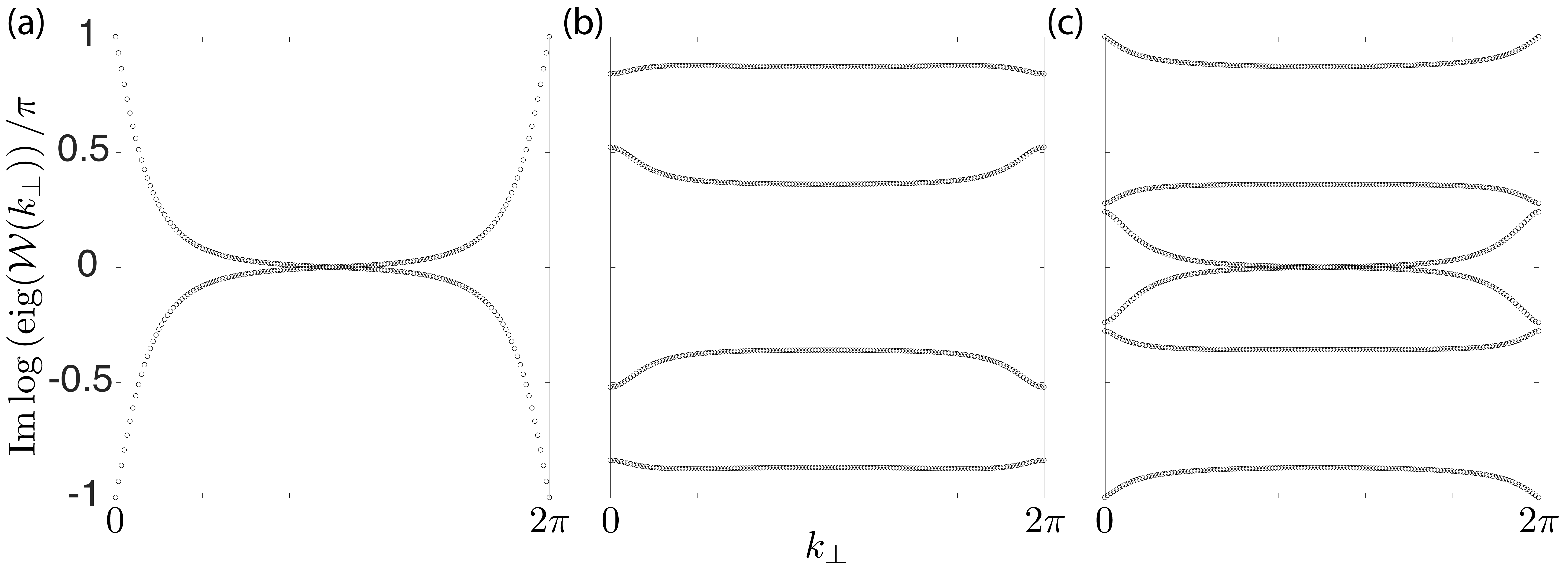}}
\caption{
Wilson loop spectra computed following the scheme described in Ref.\ \onlinecite{Song2018}, for the following set of bands in the ten-band model: (a) the two nearly flat bands at charge neutrality; (b) the lowest four bands; and (c) the six bands of (a) and (b) combined. Note that a nontrivial spectral flow, which forbids any atomic description, is found only in (a). This is consistent with the fragile nature of the band topology.
\label{fig:Wilson}
 }
\end{center}
\end{figure}

We compute this Wilson loop spectrum for our ten-band model $H(t_0,1)$ (Eq.\ \eqref{eq:10Band} in the main text), choosing $\vec b_\parallel =  \vec b_2$ and $\vec b_\perp = \vec b_1$.
The results are shown in Fig.\ \ref{fig:Wilson}, which can be studied using the extensive results derived earlier in Ref.\ \onlinecite{Ahn2018} concerning the $ C_2\mathcal T$-protected topological properties of the Wilson loops.

The Wilson loop spectrum for the two nearly flat bands is shown in Fig.\ \ref{fig:Wilson}a. We find the same nontrivial spectral flow for the nearly flat bands as in Ref.\ \onlinecite{Song2018}. This is expected, as it is quite likely that the $\mathbb Z$-valued Wilson loop invariant suggested in Ref.\ \onlinecite{Song2018} (arising from $\pi_1({\rm O}(2)) = \mathbb Z$, as was noted earlier in Refs.\ \onlinecite{Ahn2018,Slager2018}), is equivalent to the chirality obstruction we identified in Ref.\ \onlinecite{Mar2018}.  From the spectrum, one can utilize the characterization in Ref.\ \onlinecite{Ahn2018} to infer that the set of two nearly flat bands has trivial weak invariants (which are simply 1D Berry phases quantized to $0$ vs.\ $\pi$), but a nontrivial second Stiefel-Whitney (SW) class \cite{Nakahara} of $w_2 =1 \in \mathbb Z_2$. The meaning of having $w_2=1\in\mathbb Z_2$ is that there is an $O(n)$ monopole inside the BZ torus \cite{Fang2015, Ahn2018}. For two-band problems, i.e., $n=2$, our results imply the total monopole charge is half of the net chirality \cite{CastroNeto2011, He2013, Goerbig2017, Cao2018,Mar2018,Jun2018}. 
Also, we note that $w_2$ is an additive invariant with respect to band stacking, subjected to the constraint that the weak 1D invariants of the bands are trivial \cite{Ahn2018}. This constraint is satisfied by all the sets of bands we consider here.

Fig.\ \ref{fig:Wilson}b shows the spectrum obtained for the lowest four bands, which, as we have shown, are adiabatically connected to an explicit atomic insulator. Note the existence of continuous spectral gaps separating each band from the rest. Based on the characterization in Ref.\ \onlinecite{Ahn2018}, all topological invariants are trivial. This is consistent with the atomic nature of the four complementary bands.

Fig.\ \ref{fig:Wilson}c shows the spectrum for the lowest six bands, i.e., the composite bands formed by those in (a) and (b). Importantly, a continuous spectral gap is also found, similar to that observed in Ref.\ \onlinecite{Slager2018}, which is consistent with the fact that these six bands together form an atomic insulator. 
Despite its atomic nature, the characterization in Ref.\ \onlinecite{Ahn2018} indicates that the SW invariant is $w_2=1$. In fact, this follows simply from the additive nature of $w_2$. In any case, as we have demonstrated in the main text, these six bands are smoothly deformable into a strict atomic limit. This implies the proposed SW insulator in Ref.\ \onlinecite{Ahn2018} is actually atomic in nature.

More generally, we remark that our results provide a concrete physical interpretation of the 2D $C_2 \mathcal T$-protected SW invariant \cite{Ahn2018}. Recall the lowest six bands in our ten-band model corresponds to the atomic insulator $(\tau, p_z)  \oplus (\tau, p_{\pm}) \oplus (\kappa, s)$. 
We have shown that
\begin{equation}\begin{split}\label{eq:}
w_2 \left[ (\tau, p_z)  \oplus (\tau, p_{\pm}) \oplus (\kappa, s)\right] = 1.
\end{split}\end{equation}
Since the Wilson loop of the atomic insulator  $(\tau, p_z)  \oplus (\tau, p_{\pm})$ is identity in the strict atomic limit, we may conclude $w_2 \left[ (\tau, p_z)  \oplus (\tau, p_{\pm}) \right ] = 0$. 
More carefully, this can be argued as follows: First, notice that $w_2$ is well defined so long as $C_2\mathcal T$ and lattice translation symmetries are retained. Imagine breaking $C_3$ and $M_y$, such that there is no symmetry distinction between the orbitals which we originally labeled as $s$, $p_z$, $p_+$, and $p_-$. This implies $w_2 \left[ (\tau, p_z)  \oplus (\tau, p_{\pm}) \right ] = 3 w_2 \left[ (\tau, s)\right ]= w_2 \left[ (\tau, s)\right ]$. Then our claim follows as the single band problem $(\tau, s)$ is in the trivial SW class $w_2 = 0$ \cite{Ahn2018}. 

Using the additive nature of $w_2$ with respect to band stacking \cite{Fang2015, Ahn2018}, we can conclude
\begin{equation}\begin{split}\label{eq:}
1 = w_2 \left[ (\tau, p_z)  \oplus (\tau, p_{\pm}) \oplus (\kappa, s)\right] = 
w_2 \left[ (\tau, p_z)  \oplus (\tau, p_{\pm}) \right] + w_2 \left[ (\kappa, s)\right] = w_2 \left[ (\kappa, s)\right].
\end{split}\end{equation}
By definition, $(\kappa, s)$ is manifestly atomic, and hence these bands are regarded as trivial in our context. Therefore, $w_2$ only indicates (stable) mutual distinction between atomic insulators, but does not imply nontrivial band topology which forbids any atomic (i.e., product-state) description.
Lastly, we note that $w_2 [(\eta, \ell)] = 0$ for $\ell = s,p_z, p_\pm$. This is because the two honeycomb sites in each unit cell are related by $C_2 \mathcal T$ and, upon the breaking of $C_3$ and $M_y$, one can smoothly collapse the two honeycomb sites at the point-group origin while respecting the protecting symmetries for $w_2$.

\begin{figure}[h]
\begin{center}
{\includegraphics[width=0.75 \textwidth]{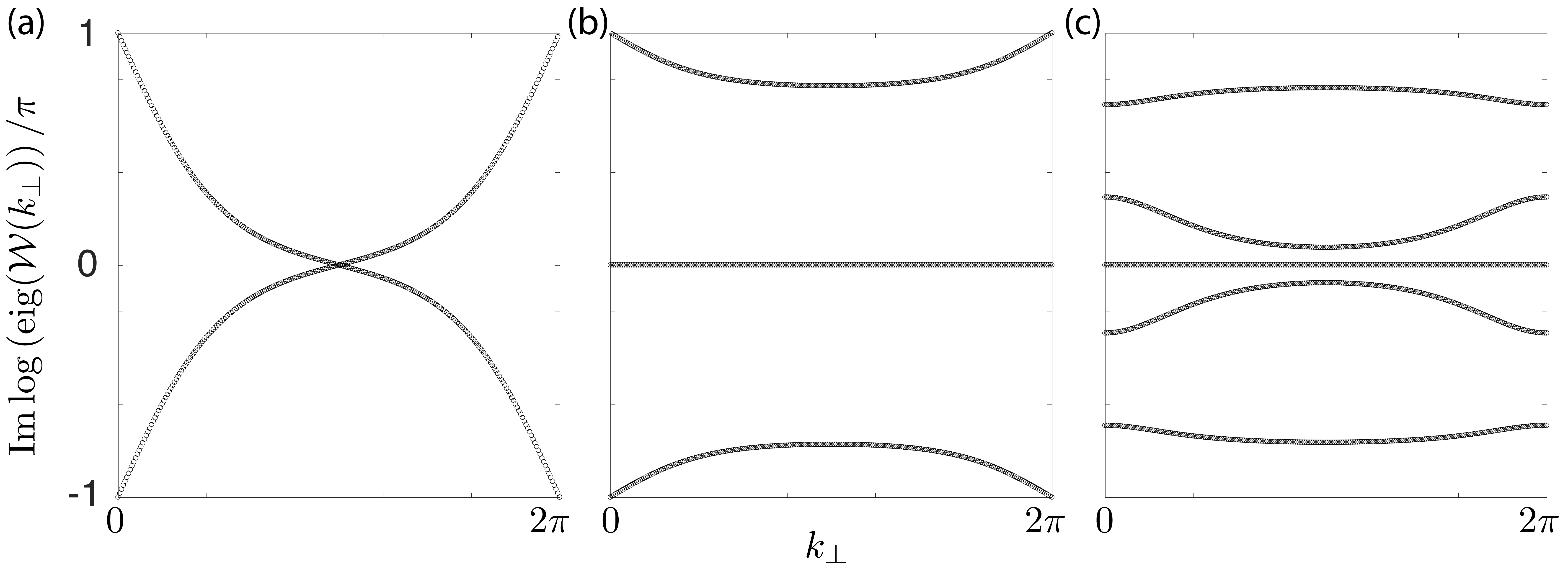}}
\caption{
Wilson loop spectra computed following the scheme described in Ref.\ \onlinecite{Song2018}, for the following set of bands in $H^{(5)}_{\mu = 0}$: (a) the two nearly flat bands at charge neutrality; (b) the lowest three bands; and (c) the five bands of (a) and (b) combined. 
\label{fig:Wilson_5Band}
 }
\end{center}
\end{figure}

With these observations in mind, one may inspect the representation-matching equations in Appendix \ref{app:SymRep} again and demand the equality of $w_2$ on the two sides. This narrows down the minimal ``fragile resolution'' of the band topology to involve $5$ bands (if the complementary bands are allowed to be topological, four-band models are possible, as we showed in Ref.\ \onlinecite{Jun2018}). 
We have already constructed a five-band fragile resolution in Appendix \ref{app:5Band}. For completeness, we also compute the Wilson loop for that model. As the Wilson loop spectrum will be completely flat in a strict atomic limit, instead of studying $H^{(5)}$ directly we consider $H^{(5)}_{\mu = 0}$ in Eq.\ \eqref{eq:Deform5Band}, whose lowest five bands have the same band topology as $H^{(5)}$  due to the persistence of band gaps as $\mu$ is varied. The results are shown in Fig.\ \ref{fig:Wilson_5Band}, which verify our earlier discussions on the relation between the $w_2$ invariant \cite{Ahn2018} and atomic insulators. In particular, the Wilson loop invariant defined in Ref.\ \onlinecite{Song2018} is trivial in Fig.\ \ref{fig:Wilson_5Band}c, which is obtained by simply appending a set of atomic bands, corresponding to $(\kappa, s)$, to the two nearly flat bands. 

\end{document}